\documentclass[11pt]{article}
\usepackage{amsmath}
\usepackage{graphicx}
\usepackage{amssymb}
\usepackage{amsfonts}
\usepackage{latexsym}
\usepackage{psfig}

\def\beq{\begin{eqnarray}}
\def\eeq{\end{eqnarray}}
\def\ln{\,\mbox{ln}\,}
\def\Det{\,\mbox{Det}\,}
\def\det{\,\mbox{det}\,}

\def\Tr{\,\mbox{Tr}\,}

\def\al{\alpha}
\def\be{\beta}

\def\ga{\gamma}
\def\de{\delta}
\def\vp{\varepsilon}
\def\ep{\epsilon}
\def\ze{\zeta}

\def\la{\lambda}
\def\na{\nabla}
\def\pa{\partial}

\def\si{\sigma}
\def\om{\omega}
\def\ph{\varphi}

\def\Ga{\Gamma}
\def\De{\Delta}
\def\La{\Lambda}

\def\Om{\Omega}

\begin{document}

\begin{center}
{\Large\sc Effective Action of Vacuum: Semiclassical Approach}
\vskip 6mm

{\bf Ilya L. Shapiro}
\vskip 2mm

{\small\sl  Departamento de F\'{\i}sica -- ICE,
Universidade Federal de Juiz de Fora }

 {\small\sl  Juiz de Fora, CEP: 36036-330, MG,  Brazil
 \footnote{Also at Tomsk State Pedagogical University,
 Russia. \ E-mail address: shapiro@fisica.ufjf.br} }

\vskip 6mm
\end{center}
\vskip 2mm


\begin{quotation}

\centerline{\Large\bf Abstract.}
\vskip 4mm

\noindent 
We present brief, to great extent pedagogical review on
renormalization in curved space-time and of some recent 
results on the derivation and better understanding of 
quantum corrections to the action of gravity. The paper 
is mainly devoted to the semiclassical approach, but we 
also discuss its importance for quantum gravity and string 
theory. 
\vskip 6mm

\tableofcontents
\pagenumbering{arabic}
\vskip 6mm

\vskip 2mm

\vskip 2mm

\\
$\,$ Conformal Anomaly.

\end{quotation}

\vskip 12mm

\section{Introduction. Classical gravity
and its applicability.} 

\qquad
 One of the most important and well understood aspects of 
 Quantum Gravity is the
 so-called semiclassical approach, where only matter fields 
 are quantized, while metric is treated as a classical 
 background. The main challenge in this area is deriving 
 the quantum corrections to the classical action of gravity. 
 Despite the form of these corrections is being, in general,
 unknown, there are strong hints that they may have numerous 
 applications in cosmology and black hole physics. Moreover, 
 depending on the role of these quantum contributions, one 
 can or not justify the importance of other branches of 
 Quantum Gravity, including string theory. In the present 
 review paper, we describe a
 recent progress in calculating and better understanding 
 quantum contributions to the effective action of gravity. 
 Let us start by presenting the standard arguments for the 
 necessity of quantum approach to gravity. 

The modern gravitational physics is mainly based on General 
Relativity (GR). Correspondingly, a standard assumption is 
that the gravity action includes the Einstein-Hilbert term. 
The existence of the nonzero cosmological constant does 
not contradict any known principle and therefore it
can be also included into the Lagrangian. So, the initial 
action has the form
\beq
S_{EH}
\,=\,-\,\frac{1}{16\pi G}\int d^4 x\sqrt{-g}\,
\left(\,R+2\La\,\right)\,.
\label{EH}
\eeq
The GR theory based on this action passed almost all known 
experimental and observational tests, however it is not 
free of unsolved problems. A serious one is, for instance, 
to explain the typical rotation curve observed in spiral 
galaxies. However, this problem can be solved by assuming 
the presence of an invisible Dark Matter - forming the halo 
in the region around the galaxy (see, e.g., the books 
\cite{Peebles,Padman-book,Dodelson,Mukhanov}). Alternatively, 
one can accept some modification in Newton's and Einstein's 
gravity laws \cite{Milgrom,Bek}, although this option 
currently looks less favorable with respect to the 
full set of existing cosmological data. Another remarkable 
example of an unexplained gravitational phenomenon is the 
problem of Pioneer anomaly \cite{Pioneer}, 
but there are still chances to explain it in the framework 
of GR or some of its modifications at the classical level. 
This can be achieved either by introducing scalar fields 
in the braneworld scenario (see, e.g., \cite{Orfeu-brane}) 
or by introducing some {\it ad hoc} curvature dependence 
to the matter and gravity Lagrangians \cite{Orfeu-fR}
\footnote{It is remarkable that modifications of 
similar form may show up due to the quantum corrections.
This issue definitely deserves a detailed study.}. 

All this shows that GR, despite its simplicity, beauty 
and efficiency, is not really a final word in the 
gravitational physics. On the top of that, GR has a 
serious conceptual problem related to the existence 
of singularities. These singularities emerge in the 
most important solutions such as spherically symmetric 
and cosmological ones. In the next two subsections we 
shortly consider these two solutions. 

\subsection{The Schwarzschild solution}

\qquad
The Schwarzschild solution corresponds to the spherical 
symmetry in the static mass distribution. In the 
simplest case of a point-like mass in the origin of the 
spherical coordinate system, the solution has the standard 
form 
\beq
ds^2\,=\,\Big(1-\frac{2GM}{r}\Big)\,dt^2
\,-\,\Big(1-\frac{2GM}{r}\Big)^{-1}\,dr^2\,-\,r^2d\Om\,,
\label{b3}
\eeq
where $d\Omega$ is the metric for the unit two-sphere.
Let us notice that if performing an expansion of the
above solution into the series in the parameter $1/r$, 
one arrives at the relativistic corrections to the 
classical gravitational potential with relativistic 
corrections
\beq
g_{00}=1+2\ph(r)\,,\qquad
\ph(r)\,=\,-\frac{GM}{r} + \frac{G^2M^2}{2r^2}+...\,\,.
\label{relativity}
\eeq

The Schwarzschild solution (\ref{b3}) contains two 
singularities: one at the gravitational radius $r_g=2GM$ 
and second at the origin $r=0$. It is well-known that 
the first singularity
is coordinate-dependent (see e.g. \cite{LL-2,Weinberg72}). 
This singularity indicates the existence of a horizon, such 
that the light signals can not propagate from the interior 
of the black hole to its exterior. The same concerns, of 
course, the massive particles, which can never escape the
interior of the black hole. However, the $r=r_g$ horizon 
looks as singularity only if it is observed from the safe 
distance. On the contrary, if the observer is changing his
coordinate system to the Kruskal one,
\beq
 u = t - r - r_g\ln \Big|\frac{r}{r_g} -1\Big|\,,\qquad
 v = t + r + r_g \ln \Big|\frac{r}{r_g} -1\Big|\,,
\label{coed}
\eeq
the metric becomes regular at $r=r_g$,
\beq
ds^2 = -\Big(1 - \frac{r_g}{r}\Big) dudv + r^2 d\Omega^2\,.
\label{edfi}
\eeq

The consistency of classical gravity is actually spoiled by the 
second singularity at the $r=0$ point. In the immediate vicinity 
of this singularity the curvature invariants grow infinitely and
therefore it can not be cured by change of coordinates. Indeed, 
the  Schwarzschild solution is valid only in the vacuum, and in 
reality one can not expect to meet point-like masses. The 
spherically symmetric solution inside the continuous 
matter does not have the $r=0$ singularity \cite{LL-2}.

The fundamental significance of the $r=0$ singularity becomes
clear after considering the phenomenon of the gravitational
collapse, which is one of possible ways of the black hole
formation (see e.g. \cite{LL-2,Frolov}). Qualitatively, 
the situation looks as follows: When the star (which has to 
be sufficiently massive) is losing its energy due to nuclear 
reactions and cools down, its size gets smaller, and the 
strength of the gravitational force on its surface increases.
This process proceeds until the star becomes very small
(a white dwarf or neutron star, depending on the initial 
state). Then, if the size of the star is large enough, 
the gravity force on the surface can become greater than 
certain limit, set up by the nuclear forces (Chandrasekhar 
or Tolman-Oppenheimer-Volkoff limits). In this case, the 
collapse of
the star will continue and eventually its radius becomes
smaller than its gravitational radius $r_g=2GM$. For an
external observer the star converts into a black hole.

The process of collapse will continue inside the 
black hole horizon (see, e.g, \cite{Frolov}). After all, 
if one assumes that the GR is valid at all scales, we 
arrive at the situation when the $r=0$ singularity becomes 
real. The matter energy density becomes infinite and so 
does the curvature. Our physical intuition tells that this 
is not a realistic situation and that some modifications 
are necessary in order to address this 
situation. The most natural option is perhaps to modify 
the action (\ref{EH}) in such a way which could prevent 
the formation of singularity at $\,r=0$.

\subsection{Standard cosmological model}

\qquad
Another important solution of GR is the one for the 
homogeneous and isotropic metric, corresponding to the 
Friedmann-Lema\^{\i}tre-Robertson-Walker (FLRW) solution. 
The most general form for the metric with these symmetries is 
\beq
ds^2\,=\,
dt^2\,-\,a^2(t)\cdot \Big(\frac{dr^2}{1-kr^2}+r^2d\Om\Big)\,,
\label{c1}
\eeq
where $r$ is the distance from some given point of the 
space, $a(t)$ is the conformal factor of the metric, and 
$k=0,1,-1$. The value of $k$ defines the curvature of 
the three-dimensional space section of the four-dimensional 
space-time manifold $M_{3+1}$. 

The cosmological model is based on the Friedmann equation
\beq
H^2\,=\,\left( \frac{{\stackrel{.} {a}}}{a}\right)
\,=\,\frac{8\pi G}{3}\,\rho - \frac{k}{a^2}\,,
\label{c5}
\eeq
where $\,H=\frac{{\stackrel{.} {a}}}{a}\,$ is a Hubble 
parameter and $\,\rho=\sum\rho_k\,$ is the total energy 
density, including contributions from different fluids, 
such as baryonic and dark 
matter, radiation, energy density of vacuum and maybe 
some additional unknown components. Furthermore, there 
are other equations, including the conservation law for 
the matter and vacuum components. In case of a single-fluid 
model we have
\beq
\frac{d\rho}{\rho+P}=-\frac{3da}{a}\,.
\label{c7}
\eeq
For the matter-dominated epoch one can assume a zero 
pressure $P=0$. The solution corresponds, in the 
realistic situations, to the expanding Universe. 
This is in perfect agreement with the observational 
data, telling us that the present-day Universe is 
expanding according to the Hubble law. The most likely 
values of the relative vacuum 
energy densities are $\Om_{vac} \approx 0.75$ for the 
cosmological constant (Dark Energy), $\Om_M \approx 0.25$ 
for the sum 
of dark and baryonic matter relative energy densities,
$\Om_{vac} \approx 0.0001$ for the radiation relative 
energy density. The nowadays Hubble constant is
$H_0 = 100\,h_0\ \, Km\,sec^{-1}\, Mpc^{-1}$, where
$h_{0}= 0.6\pm 0.1$.

Due to the expansion of the Universe there was a radiation 
dominated epoch, when the cosmological constant was not 
relevant for the expansion. In this regime one can set 
the pressure \ $P=\rho/3$ \ and then the solution has
the form \ $a\propto t^{-1/2}$. In the limit \ $t\to 0$ \
this leads to the coordinate-independent singularity.
Qualitatively the situation is quite similar to the one 
in the spherically symmetric solution. The classical 
solutions for the space-time manifold possess
singularities and one would think whether modification  
of the gravitational equations are requested to cure
this disease. 

\subsection{
Planck units, fundamental scale and quantum effects}

\qquad
As we have just seen, the two most realistic solutions of 
GR end up with singularities. According to the dimensional 
arguments, in the regions close to singularities some 
quantum effects may be relevant. In a sufficiently close 
vicinity of the singularities one meets energy densities 
and curvature tensor components of a Planck order of 
magnitude. 

The idea of the Planck units is the following. There are 
three fundamental constants in nature: the speed of light 
in vacuum $c$, the Planck constant $\hbar$ and the 
Newton constant $G$,
\beq
c=3 \cdot 10^{10}\,cm/sec\,,\,\,\,\,\,\,\,\,\,\,\,
{\hbar}=1.054\cdot 10^{-27}\,erg\cdot sec
\,,\,\,\,\,\,\,\,\,\,\,\,
G = 6.67\cdot 10^{-8}\,\frac{cm^3}{sec^2\, g}\,.
\label{planck 1}
\eeq 
It turns out that they can be used to construct the 
fundamental quantities with dimension of length 
\ $l_P$, time $t_P$ and mass $M_P$:
\beq
l_P &=&G^{1/2}\,{\hbar}^{1/2}\,c^{-3/2}\,\approx
\,1.4 \cdot 10^{-33}\,cm;
\nonumber
\\
t_P &=&G^{1/2}\,{\hbar}^{1/2}\,c^{-5/2}\,
\approx \,0.7\cdot 10^{-43}\,sec;
\nonumber
\\
M_P &=&G^{-1/2}\,{\hbar}^{1/2}\,c^{1/2}
\,\approx\, 0.2\cdot 10^{-5}\,g\,.
\label{planck 2}
\eeq

These fundamental units can be interpreted in different 
ways. Let us start with the particle physics, where people 
use to put $c={\hbar}=1$ and measure everything in $GeV$. 
Of course, in the everyday life this is not very nice,
since you have to schedule the meeting with your friend
``about $10^{27}\,GeV^{-1}$ from now''. It is actually 
the same thing, but ``about $15$ minutes from now'' will 
be, perhaps, better appreciated. However, in a specific 
area of particle physics there are no factors like 
$10^{27}$ and the $GeV$ units are very useful.
One can also measure the Newton constant in $GeV$,
according to (\ref{planck 2}) we have $G=1/M_P^2$.
Of course, in these units $t_P=l_P=1/M_P$. The numerical 
value is about $M_P=1.22\times 10^{19}\,GeV$.

One can suppose that the existence of fundamental
units indicates the presence of some fundamental physics. 
Since the quantities (\ref{planck 2}) involve, simultaneously,
$c,{\hbar}$ and $G$, we assume that the fundamental 
scale corresponds to some relativistic and simultaneously
quantum, gravitational physics. One can think, for example, 
that due to the quantum effects, in the corresponding regions 
of the space-time manifold the gravitational theory has to 
be modified. The corrections may come from quantum matter, 
quantum gravity, supergravity, from the superstring theory, 
or from some yet unknown theory. In 
any case we have to assume the universal nature of these
corrections. This means the 
gravitational action should be actually different from eq. 
(\ref{EH}) everywhere, not only in the vicinity of the 
singularities. And, since GR is a very successful theory, 
the first constraint is that, far from singularities, the 
effect of extra terms must be weak. 

The dimensional analysis can not tell us an exact form 
of extra terms in the gravitational action. Also, we do 
not have certainty in what new kind of physics may be 
relevant for deriving these additional terms. Let us 
outline the following three main options: 

{\it i)} Gravity and matter fields must be quantized 
at the Planck scale.

{\it ii)} Matter fields should be quantized, while 
gravity is an intrinsically classic interaction and 
hence should not be quantized at all. This looks like 
a reasonable option because, already in GR, gravity is 
different from other fields. 

{\it iii)} Neither gravity nor matter fields should be 
quantized, both below and above the Planck scale. All we 
know as ``fundamental interactions'' emerge as effective 
low-energy effects of some unknown, really fundamental 
object, which must be, indeed, quantized at the Planck 
scale.

Let us start with a very brief description of the last 
paradigm {\it iii)} which, of course, corresponds to the 
(super)string theory \cite{GSW}. In this theory the 
gravitational action is the low energy effective action 
of the background fields (metric, torsion, dilaton and 
their superpartners) of the really fundamental quantum 
object, that is a superstring. This effective action 
can be expanded into 
power series in the dimensional parameter $\alpha^\prime$. 
Einstein-Hilbert action corresponds to the lowest-order 
term of this expansion. Beyond this order one meets an 
infinite set of higher derivative terms. In any finite 
order in $\alpha^\prime$ these terms do suffer from a 
parametrization ambiguity. The origin of this ambiguity 
is that all fields, including the space-time metric, are 
nothing but external parameters for the quantum string. 
The reparametrization of these 
parameters does not spoil the consistency of the quantum  
theory of string \cite{zwei}. However, the physical effects 
which follow from the higher order corrections do depend 
essentially on such a reparametrization \cite{marot}. 
On top of that, there are other ambiguities, for example 
those related to the compactification of extra dimensions. 
Finally, although the superstring theory is mathematically 
consistent, until now there is no clear way to extract 
falcifiable physical predictions out of it. Therefore, 
if we are interested in studying quantum corrections to 
General Relativity, it is worthwhile not to restrict our 
attention on the option {\it iii)}, but also look 
somewhere else. 

In the rest of our review we shall mainly deal with option  
{\it ii)} and just comment on the option {\it i)}
now and, once again, in the special section 8. Of course, 
option {\it i)} looks more general, more fundamental and 
therefore more attractive. Furthermore, there is a standard 
argument in favor of quantizing gravity, related to a 
quantum mechanical inconsistency of the semiclassical 
approach  {\it ii)} (see, e.g., \cite{Wald} and references 
therein). 
These inconsistency, however, may be seen not as an 
argument in favor of quantizing gravity, but as a 
certain indication to the change of quantum mechanical
laws, at least in the vicinity of singularities where 
the discrepancy mentioned above can be observed. 

The main disadvantage of quantizing gravity is the 
known difficulty in formulating a consistent version of 
perturbative quantum field theory for the gravitational 
field. The attempts of quantizing gravity started long 
ago and at some point it became clear that the quantum
theory based on general relativity is not renormalizable 
\cite{hove,dene}. Alternative theories with higher 
derivatives may be renormalizable \cite{Stelle} or 
even superrenormalizable \cite{highderi} but they 
have unphysical components, called massive ghosts, 
in the spectrum and, at least 
if being treated as usual quantum field theories on 
flat background, this may lead to the violation of 
unitarity \cite{Stelle}. Indeed, there is a chance 
that the quantum corrections will make such theories
unitary \cite{tomb,antomb}, but the verification of 
this is, for a moment, beyond our possibilities 
\cite{johnston}.

Is it true that string theory is a real thing? Is it 
true that gravity should be quantized? We do not know 
the answers to these questions at the moment and it is 
not clear whether we will find them out soon. On the 
other hand, there are some certain and safe things and 
all of them concern the option {\it ii)} of our list. 
The success of the Standard Model of particle physics 
(SM) shows the correctness of the 
quantum field theory (QFT) description of the particles 
interactions. One of the most important aspects of QFT is 
the complicated vacuum structure which implies vacuum 
polarization,  the possibility of particle creation and, 
in general, relevant effects of the virtual loops of 
matter fields. The vacuum quantum effect of matter 
fields can affect the gravitational action and, in 
principle, can change a situation with the fundamental 
problems of classical gravity which were 
discussed above. It is important to keep in mind that 
these quantum effects do not correspond to some 
qualitatively new physics which may exist or not. Much 
on the contrary, they represent a relatively well known 
physics, considered in a more complicated environment, 
that is in an external gravitational field. 

Looking from this perspective, 
the most natural question is: -- in which way vacuum quantum 
effects of matter fields, e.g., in the framework of SM or 
GUT's, do contribute to the gravitational action? 
The complete answer to the last question is unknown. 
The main purpose of this review is a survey of known 
corrections from quantum matter fields and their most 
important physical implications. At the end, we shall 
also discuss how the semiclassical results can, in 
principle, affect the necessity of a more complete 
quantum gravitational theory.

\section{The semiclassical approach: choice of the action}

\qquad 
The introduction to standard QFT in curved space can be 
found in the books
\cite{BDW-65,GMM, birdav,book,Frolov,MukhWitn}. In this 
and consequent sections we just review some fundamental 
aspects of the theory. 

\subsection{Classical action}

\qquad 
The first step is to formulate quantum theory of matter 
on classical curved background, that means, at the first 
place, to define the classical actions for matter fields 
and for gravity. One can formulate these actions in 
infinitely many ways, so let us discuss only the simplest 
and most natural versions of the theory which provide 
the consistency at quantum level. Following this line, 
we impose the principles of locality and general 
covariance for both matter and gravity sectors. Furthermore, 
in order to preserve the fundamental features of the original 
flat-space theory, one has to require the symmetries (gauge 
invariance, at the first place) which take place in flat 
space-time, to hold for the more general theory in 
curved space-time. Even after that the number of possible 
terms in the action is unbounded, so we need some 
additional restriction. The natural requirement for the 
theory in curved space is renormalizability and, as we have 
already mentioned, simplicity. As we shall see in what 
follows, these two conditions can be satisfied even if 
we forbid new parameters with the inverse-mass dimension. 
This set of conditions enables one to construct the 
consistent quantum theory of matter fields on the classical 
gravitational background. Following the mentioned three 
principles (locality, covariance and restricted dimension),
the form of the action is fixed except the values of some 
new parameters which remain arbitrary. The procedure which 
we have described above, leads to the so-called non-minimal
actions.

Besides the nonminimal scheme for constructing the actions 
in curved space, described above, there is also a more simple, 
minimal one. According to it the partial derivatives 
$\pa_\mu$ are substituted by the covariant ones 
$\nabla_\mu$, the flat metric $\eta_{\mu\nu}$ by 
$g_{\mu\nu}$ and the volume element $d^4x$ by the 
covariant expression $\,d^4x\sqrt{-g}$. Let us compare 
the results of the two approaches. For the free scalar 
field the nonminimal generalization of the action is
\beq
S_{scal}\,=\,\frac12\,\int d^4x\,\sqrt{-g}\left\{\,
g^{\mu\nu}\,\pa_\mu\ph\,\pa_\nu\ph
\,+\,m^2\,\ph^2\,+\,\xi\,\ph^2\,R
\,\right\}\,,
\label{scalar1}
\eeq
If compared to the flat-space theory,
the action (\ref{scalar1}) involves a new dimensionless 
quantity $\xi$ which is called nonminimal parameter. 
The minimal version has $\xi=0$. It is easy to check 
that the $\xi$-dependent term is the unique one which 
is admitted in the scalar sector by the three 
principles imposed above. In the case of the 
multi-scalar theory $S[\ph^i]$ the nonminimal term 
is $\,\int d^4x\sqrt{-g}\xi_{ij}\ph^i\ph^j\,R$. 
The generalization to the case of a charged scalar is 
obvious. The non-minimal term in the action 
(\ref{scalar1}) looks as a kind of modification 
of the massive term, despite there is a very essential 
difference between them. We shall discuss this difference 
in the subsection 2.3. 

It is very important that the mentioned three principles 
enable one to introduce just a finite number of 
matter-independent, purely vacuum terms. These terms 
represent a qualitatively new element compared to the 
flat-space theory. The most general action of vacuum,
according to our principles, is as follows:
\beq
S_{vac} &=& S_{EH}\,+\,S_{HD}\,,
\label{vacuum}
\eeq
where $\,S_{EH}\,$ is the Einstein-Hilbert action with 
the cosmological constant (\ref{EH}) and
\beq
S_{HD} &=& \int d^4x \sqrt{-g}
\left\{a_1C^2+a_2E+a_3{\Box}R+a_4R^2 \right\},
\label{HD}
\eeq
where $\,C^2=R_{\mu\nu\al\be}^2 - 2 R_{\al\be}^2 + (1/3)\,R^2\,$
is the square of the Weyl tensor and 
$\,E = R_{\mu\nu\al\be}^2 - 4 R_{\al\be}^2 + R^2\,$
is the integrand of the Gauss-Bonnet topological term. Let us 
remark that the presence of higher derivative terms is 
unavoidable is one wants have renormalizable theory. The same 
concerns the cosmological term, especially in case matter fields
are massive. 

The higher derivative terms in (\ref{vacuum}) are not quantum 
corrections, they should be introduced already at the classical 
level. The reason why they are not observed in the gravitational 
experiments is that the lower derivative Einstein-Hilbert term 
has the coefficient $\,{1}/{G}=M_P^2$. Independent on whether 
we quantize gravity or not it is useful to use the language of 
Feynman diagrams. For example, the Newton law is the consequence 
of one-graviton exchange between the two masses. The higher 
derivative terms produce an additional particle similar to 
the graviton, but with the mass of the order 
$\,m_2\sim M_P/\sqrt{|a_1|}\,$ or 
$\,m_0 \sim M_P/\sqrt{|a_4|}\,$, depending on the sector 
(spin-two or spin zero) of the propagator \cite{Stelle-78}. 
The exchange of these new particles produce modifications of 
the Newton law, but these corrections are small for the 
small values of transfered momenta, for the propagator can 
be presented, e.g., as 
\beq
\frac{1}{k^2+m_2^2}=\frac{1}{m_2^2}
\,\left(1 - \frac{k^2}{m_2^2} + \frac{k^4}{m_2^4} - \,\, ...
\right)\,.
\label{classic_decoupling}
\eeq
It is obvious 
that, in order to have relevant impact of the higher derivative
terms on the Newton law one needs the energy of graviton 
comparable to the Planck mass. Of course, the difference 
between the Planck scale and the present-day astrophysical 
scale is huge. As a consequence the higher derivative terms 
are irrelevant not only in the Solar System but also in most 
of the astrophysical objects, maybe except the black 
holes and sources of the gamma-ray bursts. 
In cosmology the higher derivative terms  can be relevant 
only in the very early stage of the evolution of the Universe, 
e.g. in the Starobinsky inflation \cite{star}.    

In order to complete the story, let us notice that the 
covariant, gauge invariant and local actions for fermion 
and vector fields are minimal, because no non-minimal 
terms are alrebraically possible. For the spinors we have
\beq
S_{fermion} \,=\, i\,\int d^4x\sqrt{-g}\,\left(\,
{\bar \psi}\,\ga^\al \,{\na}_\al\psi 
 -  im\, {\bar \psi} \psi\,  \right)\,,
\label{dirac 1}
\eeq
where $\ga^\mu$ and $\na_\mu$ are gamma-matrix and covariant
derivative of the spinor in curved space-time. For the gauge 
fields the action is
\beq
S_{vector} \,=\,-\,
\frac14 \,\int d^4x\sqrt{-g}\,G^a_{\mu\nu}\,G^{a\,\mu\nu}\,,
\quad \mbox{where} \quad 
G^a_{\mu\nu}=\pa_\mu A^a_\nu-\pa_\nu A^a_\mu
+ g f^{abc}A^b_\mu A^c_\nu\,.
\label{vacuum 5}
\eeq

The interactions between the matter fields (gauge, Yukawa and 
self-scalar ones) are defined through the minimal procedure, 
because there are no nonminimal extensions compatible with the 
principles declared above. Finally, the vacuum action 
has universal form (\ref{vacuum}), independent on the choice 
of the fields and their interactions.

The last remark concerns a very important feature of the 
massless version of the matter fields and vacuum actions 
formulated above. In the scalar case (\ref{scalar1}) the 
value $\xi=1/6$ corresponds to the local conformal symmetry, 
that means the equations of motion of the theory do not 
change under the transformation 
\beq
g_{\mu\nu} \to g_{\mu\nu}^\prime = g_{\mu\nu} \, e^{2\si(x)}
\,\quad
\ph \to \ph^\prime = \ph \, e^{-\si(x)}\,.
\label{trans}
\eeq
For spinor and vector fields in curved space the conformal 
transformations have the form
\beq
A_\mu \to A_\mu^\prime = A_\mu
\,\qquad
\psi \to \psi^\prime = \ph \, e^{-3\si(x)/2}\,.
\nonumber
\eeq
The local conformal symmetry of the free actions for 
these fields does 
not require anything but the masslessness. 
The form of the Noether identity corresponding to the 
symmetry under the local conformal transformation is 
\beq
\Big[
\,2\,g_{\mu\nu}\,\frac{\de}{\de g_{\mu\nu}}
+\,\sum_i k_i\,\Phi_i\,\frac{\de}{\de \Phi_i}\,\Big]
\,S(g_{\mu\nu},\,\Phi_i)\,=\,0\,,
\label{Noether}
\eeq
where $\,k_i=\left(-1,\,-3/2,\,0\right)\,$ are the 
conformal weights of the matter fields 
$\,\Phi_i=\left(\ph,\,\psi,\,A_\mu\right)$. 

In the vacuum action (\ref{vacuum}), only the higher 
derivative terms may satisfy the Noether identity 
(\ref{Noether}). The condition of conformal symmetry 
for the action (\ref{HD}) has the form \ $a_4=0$. 
Let us notice that only the Weyl term 
$\int \sqrt{-g}C^2$ is a real conformal invariant, 
while the topological $\,\int \sqrt{-g}E\,$ and 
surface $\,\int \sqrt{-g}\Box R\,$ 
terms in the vacuum action do change under the metric 
transformation (\ref{trans}). However, since they 
do not contribute to the equations of motion, the 
conformal Noether identity is satisfied for these 
terms. 

It is a custom (or, better say, tradition) to consider 
the higher derivative terms (\ref{HD}) as producing the 
energy-momentum tensor $\,T^\nu_\mu\,$ of vacuum. Then 
the Noether identity (\ref{Noether}) can be seen as the 
condition of a zero trace,  $\,T^\mu_\mu=0$. 
In the consequent sections we shall explore the important 
role of the violation of this condition at the quantum 
level. This violation is behind the most important 
applications of QFT in curved space-time, such as 
Hawking radiation \cite{Hawking75,ChrFull} and 
Starobinsky model of inflation \cite{star}.

\subsection{Note on inflation and the natural candidate 
for being the inflaton}

\qquad
Inflation is one of the most natural applications of 
Quantum Field Theory in curved space-time, because it
occurs at very high energies, where the quantum effects 
may be indeed relevant. An increasing interest to 
inflation emerged after the paper by Guth \cite{Guth},
where he noticed that the period of extremely fast
expansion of the Universe possesses the following two 
properties: \ {\it a)} Solves numerous problems of the 
standard cosmological scenario \cite{Peebles,Kolb} 
such as the ones of flatness, horizon, monopole etc;  
\ {\it b)} Can be results of the Spontaneous Symmetry 
Breaking in the Standard Model of particle physics. In other 
words, the naturality of inflation is directly 
related to the natural origin of inflaton, which is 
nothing else but the Higgs field in the original 
proposal \cite{Guth}. Let us notice, by passing, 
that the original proposal for the SM-based inflation 
may be rescued from the known difficulties \cite{Linde} by 
the non-minimal generalization of the Higgs field, that 
is by adding the $\,\xi RH^\dagger H$-term to the 
Higgs potential \cite{Shaposh}. Needless to say this 
step is also the most natural one in view of the critical 
importance of a nonminimal scalar-curvature coupling for 
the renormalizability of the theory. If the non-minimally 
interacted to gravity Higgs field really satisfies all 
tests of phenomenological cosmology, it can be seen as 
the first candidate to be inflaton!
  
Another possibility to have natural inflation is related to 
the account of the vacuum quantum effects, that is in the 
framework of original Starobinsky model \cite{star} or 
its modified version \cite{anju,Shocom,asta}. We shall 
discuss the backgrounds of these models in section 5.

\subsection{Spontaneous Symmetry Breaking and induced gravity}

\qquad 
Before starting the consideration of the loop effects, it 
is worthwhile to spend some time and discuss one important 
aspect of the tree-level QFT in curved space-time. It is 
a classical issue in the 
sense it does not depend on the loop corrections. On the 
other hand, we shall observe here one typically quantum 
property, that is the presence of non-localities. Even 
more important is that we shall meet here a very important 
aspect of QFT in curved space, that is the possibility to 
induce the gravitational action. 
Let us remember that 
the QFT models of our main concern are the Minimal Standard 
Model of particle physics and also its extensions and 
generalizations, such as GUT's (Grand Unification Theories). 
The definition of vacuum is a very important element of 
these theories and it is usually performed through the 
Spontaneous Symmetry Breaking (SSB) and the Higgs mechanism. 
So, let us check, following  \cite{sponta}, what is the impact 
of an external gravitational field here.

It is well known that the SSB leads to the induced
cosmological constant \cite{zeld-67} and in general to 
the induced gravity \cite{Sakharov} (see \cite{adler} for
a general review of induced gravity). Here we shall arrive  
at the induced action of gravity in a most natural way 
and also observe the emergence of the nonminimal terms.
We associate the scalar (\ref{scalar1}) with the Higgs 
field. Consider the classical potential 
\beq
V(\ph)\,=\,-\,\mu_0^2\,\ph^*\ph\,+\,\la(\ph^*\ph)^2
\,-\,\xi\,R\,\ph^*\ph\,.
\label{pot}
\eeq
The VEV for the scalar field is defined as a solution
of the equation of motion
\beq
-\,{\Box}v\,+\,\mu_0^2\,v\,+\,\xi R\,v\,-\,2\la v^3\,=\,0\,.
\label{SSB 2}
\eeq
If the interaction between scalar and metric is
minimal $\,\xi=0$, the SSB is standard and simple, because
the vacuum solution of the last equation is constant
\beq
v_0^2\,=\,\frac{\mu_0^2}{2\,\la}\,\,,
\label{SSB 3}
\eeq
where we have introduced a special notation
$\,v_0\,$ for the case of a minimal interaction, in order
to distinguish it from the solution $\,v\,$ of the general
equation (\ref{SSB 2}). The consistency of the QFT
in curved space requires the non-minimal interaction such 
that $\,\xi\neq 0$. Then, for a general case of a non-constant
scalar curvature one meets, instead of Eq. (\ref{SSB 3}),
another solution $\,v(x)\neq \mbox{const}$. The derivatives 
of $\,v\,$ can not be ignored and, therefore,
the solution for the VEV can not be obtained in a closed and
simple form.

For the slowly varying curvature one can use 
the eq. (\ref{SSB 3}) as the zero-order approximation. The 
solution of (\ref{SSB 2}) can be found in the form 
of the power series in $\,\xi$,
\beq
v(x)\,=\,v_0\,+\,v_1(x)\,+\,v_2(x)\,+\,...\,.
\label{SSB 4}
\eeq
For the first order term $\,v_1(x)\,$ there is equation
\beq
-\,{\Box}v_1\,+\,\mu^2v_1\,+\,\xi R\,v_0
\,-\,6\la v_0^2\,v_1\,=\,0\,.
\label{SSB 5}
\eeq
The solution has the form
\beq
v_1\,=\,\frac{\xi\,v_0}{{\Box}-\mu^2+6\la v_0^2}\, R
\,=\,\frac{\xi\, v_0}{{\Box}\,+\,4\la v_0^2}\,R\,,
\label{SSB 6}
\eeq
where we used (\ref{SSB 3}). In a similar way, we find
\beq
v_2\,=\,\frac{\xi^2\,v_0}{\Box + 4\la v_0^2}\,
R\,\, \frac{1}{\Box + 4\la v_0^2}\,R
\,-\,\frac{6\,\la\,\xi^2\,v^3_0}{\Box + 4\la v_0^2}\,
\Big(\,\frac{1}{\Box + 4\,\la v_0^2}\,R\,\Big)^2\,,
\label{SSB 7}
\eeq
One can continue the expansion of $\,v\,$ to 
any desirable order, deriving $\,v_3$, $\,v_4\,$ etc. 

Some observations are in order. Different from the usual 
SSB case, here the VEV of the scalar field is not a constant. 
Instead, it varies due to the variable curvature. Of course, 
this variation is completely negligible for the particle 
physics due to the extremely small value of curvature compared 
to any other dimensional quantity such as, e.g., $v_0$. 
Hence the impact of the gravitational interaction on the 
particle physics applications is irrelevant. 

In the gravitational applications, however, the effect of 
SSB and nonlocalities is nontrivial. If one replace 
the SSB solution $\,v(x)\,$ back into the 
the action of the scalar field, the following result for 
the induced low-energy action of vacuum will follow:
\beq
S_{ind}
\,=\,\int d^4x\sqrt{-g}\,\Big\{\,g^{\mu\nu}
\,\pa_\mu v \,\pa_\nu v \,+\,(\mu_0^2+\xi R)\,v^2
\,-\,\la\,v^4\,\Big\}\,.
\label{SSB 80}
\eeq
In the lowest order of power expansion (\ref{SSB 5}) the 
VEV of the scalar is $\,v \approx v_0\,$ and we meet 
the custom form of the induced Einstein and cosmological 
constant terms. 
The induced values of the Newton and cosmological 
constants correspond to
\beq
\frac{1}{16\pi\,G_{ind}}\,=\,-\,\xi\,v_0^2\,,\qquad
\frac{\La_{ind}}{8\pi\,G_{ind}}\,=\,-\,\mu_0^2\,v_0^2
\,+\,\la\,v_0^4\,=\,-\,\la\,v_0^4\,.
\label{ind EH}
\eeq
As we see, at the $v_0$ level, the low-energy induced 
action of gravity, due to the SSB, has the same form as 
the classical vacuum Einstein-Hilbert action with the 
cosmological constant. The difference between the two 
actions is that the constants $G$ and $\La$ of the 
vacuum action (\ref{EH}) are independent parameters
while the induced quantities (\ref{ind EH}) depend
on the VEV and $\xi$. Other symmetries breaking may 
result in additional contributions to the induced 
action of gravity. In the simplest formulation, the 
induced quantities (\ref{ind EH}) have to be 
summed up with the  vacuum quantities, that are 
independent parameters of the action (\ref{EH})
\footnote{In the
more extreme version of the theory \cite{adler} the initial 
gravitational action is absent and all gravity is induced. 
This is indeed a very interesting and fruitful idea, but 
it goes beyond the purpose of the present review.}. 
The observed quantities are the sums
\beq
\frac{1}{G_{obs}}=\frac{1}{G}+\frac{1}{G_{ind}}\,,
\qquad
\frac{\La_{obs}}{G_{obs}}
=\frac{\La}{G}+\frac{\La_{ind}}{G_{ind}}\,.
\label{sums}
\eeq
It is interesting that both vacuum and induced quantities 
become subjects of renormalization at the quantum level. 

If one compares the magnitudes of the vacuum and observed 
quantities, it is easy to notice that the effect of $G_{ind}$ 
is quite small. For instance, consider $G_{ind}$ in the 
framework of the Minimal Standard Model (MSM) or GUT. 
In case of MSM, $v_0$ is 
about $250\,GeV$ and therefore the induced quantity of 
\ $(G_{ind})^{-1}$ \ is about 32 orders of magnitude 
smaller than the observed one \ $(G_{obs})^{-1}$.
The same difference is about 5 orders of magnitude 
in the GUT case. 
On the contrary, the induced value of the cosmological 
constant is enormous compared to the observed quantity, 
leading to the famous, important and mysterious 
cosmological constant problem. 
The induced cosmological term is supposed to almost 
cancel with its vacuum counterpart. The precision of the 
required cancelation is 55 orders of magnitude in the 
framework of the SM and even much more than that in GUT´s 
and other generalizations of the SM. The nowaday physics 
can not explain the origin of this fine-tuning and this
is one of the most difficult conceptual problems. 
One can find an extensive discussion of this 
problem in \cite{weinberg89} and also, from the point 
of view of renormalization theory, in \cite{nova}. 

Beyond the lowest order one has to take into account 
the next terms in the expansion (\ref{SSB 4}) and thus 
to account for the space and time dependence of the 
curvature scalar. It is remarkable that, along with the
conventional cosmological constant and Einstein-Hilbert 
term, here we meet also an infinite series of non-local 
additional expressions due 
to non-localities in (\ref{SSB 4}). For example, in the 
second order, by performing an expansion in the powers of 
the curvature tensor,  we obtain much more complicated 
form of the induced gravity action
\beq
S_{ind}
\,=\,\int d^4x\sqrt{-g}\,\Big\{-\,v_1{\Box}v_1\,+\,
\mu^2\,(v_0^2+2v_0v_1+2v_0v_2+v_1^2)
\nonumber
\\
-\,\la\,(v_0^4+4v_0^3v_1+4v_0^3v_2+6v_0^2v_1^2)
\,+\,\xi R\,(v_0^2+2v_0v_1)\,\Big\}\,+\,{\cal O}(R^3)\,.
\label{SSB 8}
\eeq
Now, using the equation (\ref{SSB 5}), after a small
algebra we arrive at the following form of the action 
of induced gravity
\beq
S_{ind} &=& \int d^4x\sqrt{-g}\,\big\{
\,\la v_0^4\,+\,\xi Rv_0^2\big\}
\label{SSB 9}
\\
&+&   \xi^2\,v_0^2\,\int d^4x\sqrt{-g(x)}
\,\int d^4y\sqrt{-g(y)}\,\,
R(x)\,\Big(\frac{1}{{\Box}
+4\,\la v_0^2}\Big)_{x,y}\,R(y)\,\,+\,...\,,
\nonumber
\eeq
where the dots stand for the third, fourth and higher 
order terms in the scalar curvature $R$ with the 
corresponding insertions of the Green functions. 
Besides the usual local terms, the last (in fact, 
more precise) version of the induced tree-level
gravitational action includes an infinite set of the
non-local terms due to the specific non-constant VEV 
of the scalar field (\ref{SSB 5}) - (\ref{SSB 7}). 
The appearance
of the non-local terms in the induced action (\ref{SSB 9})
is remarkable, also, for other reasons. Although the
coefficients of these terms are very small compared to the
vacuum Einstein-Hilbert term, the non-localities do not
mix with the local terms and, in principle, can lead to
some physical effects. If considering the low energy SSB
phenomena in the framework of the SM, the non-local 
terms are irrelevant at low energies due to the large 
value of the mass term $4\la v_0^2$. But, if we assume 
that there is an extremely light
scalar (e.g. quintessence), whose mass is of the order
of the Hubble parameter and which has a potential admitting
a SSB, then the non-localities may become relevant and
in particular lead to observable consequences. This part of 
the story has not been explored so far. Hence, in the next
sections we will not discuss it and instead concentrate 
on the quantum one-loop corrections to the induced action 
(\ref{SSB 9}).

\section{Effective Action and Renormalization}

\qquad
Here we start to deal with the main subject of our 
review. At the quantum level the classical action of vacuum 
(\ref{vacuum}) is replaced by the Effective Action (EA) \
$\,\Ga [g_{\mu\nu}]$, which may be defined via path integral
\beq
e^{i\Ga [g_{\mu\nu}]}=\int {\cal D}\phi\,
e^{iS [\phi;\,g_{\mu\nu}]}\,.
\label{EA}
\eeq
Here $\phi$ is the set of all matter fields and gauge ghosts,
 $\,{\cal D}\phi\,$ is the covariant measure of functional 
integration. The classical action \ $S[\phi;\,g_{\mu\nu}]$ 
includes matter fields, interactions between these fields, 
it depends on the metric (which plays the role of external 
parameter) and also the classical action of vacuum 
(\ref{vacuum}). In the case when the background matter 
fields are present, the expression (\ref{EA}) should be 
generalized in a standard way \cite{book}. Here we shall 
mainly deal with the effective action of vacuum and therefore
consider the purely metric background. 

The effective action of gravity $\Ga [g_{\mu\nu}]$ admits a loop expansion
\cite{book}
\beq
\Ga [g_{\mu\nu}] = S_{vac}[g_{\mu\nu}]
+ {\bar \Ga}^{(1)}+ {\bar \Ga}^{(2)}+ {\bar \Ga}^{(3)}+...\,,
\label{loop}
\eeq
The simplest and usually most important 1-loop contribution 
is given by the expression
\beq
{\bar \Ga}^{(1)} = \frac{i}{2}\,\Tr \ln\,\hat{H}\,,
\label{1-loop}
\eeq
\beq
\mbox{where}\qquad
\left.
\hat{H}\,=\,\hat{H}(x,y)\,=\,\frac12\,
\frac{\de^2\,S[\phi,\,g_{\mu\nu}]}{\de \phi(x)\,\de \phi(y)}
\right|_{\phi=0}
\label{bilinear}
\eeq
is the bilinear in quantum fields part of the classical action.

The covariance of effective action can be, presumably, established using 
the same methods which are used for other gauge interactions 
\cite{vortyu,gomwein}. Furthermore, as we shall discuss later 
on, there are explicitly covariant calculational methods. The 
principal known approaches to calculate quantum corrections 
are listed below.

\subsection
{Possible and impossible form of quantum corrections}

\qquad
It is remarkable that already at this level one can make 
some essential statements about possible and impossible 
forms of quantum corrections. The effective action 
\ $\Ga [g_{\mu\nu}]$ \
is a well-defined diffeomorphism invariant quantity. As a 
consequence, \ $\Ga [g_{\mu\nu}]$ \ can not include odd 
powers of the metric derivatives. Let us emphasize that 
this property is not related to the perturbative expansion 
and is valid independent on whether the effective actionis a local functional 
of the metric (indeed, it is nonlocal, as we shall see soon). 
This important property of effective actionholds for {\it any} particular 
metrics, including the cosmological one. 

In case one detects, someday, the odd-power behavior in 
the gravitational solutions, this would be an indication 
to a certain ``new physics'', e.g. quintessence, extra 
dimensions, branes etc. However, it can not be a vacuum 
quantum effect of known fields on purely metric background.
An interesting application to cosmology is that the quantum 
corrections to the cosmological constant in the late universe, 
without scalar fields, may start from $H^2$ (here $H$ is the 
Hubble parameter in the late Universe), but not from 
$H$, because this would mean an odd metric derivative. In
particular, this rules out the hypothetical QCD contributions 
to the vacuum energy suggested in \cite{schut}. 

One can go further and make even stronger affirmative. 
Which kind of fundamental physics may be relevant for 
the possible scale dependence of the vacuum energy? 
Let us notice the relation
\beq
\rho_\La (observable) \,\, \propto \,\, 
H_0^2 M_{Planck}^2\,,
\label{CC-H2}
\eeq
where $H_0$ is the present-day Hubble parameter. The 
origin of this relation is the recent astronomical 
data concerning the energy balance of the universe
which is dominated by the ($\sim 75\%$) Dark Energy
(see, e.g., \cite{Sahni}). 
The energy 
of vacuum has mass dimension four and if the 
$\,{\cal O}(H^2_0)\,$ term is absent, the effect of 
quantum corrections can be only 
$\,{\cal O}(H^4_0) \sim 10^{-168}\,GeV^4$, that is 
negligible compared to the critical density of the 
universe $\rho_c \sim 10^{-48}\,GeV^4$. Hence we can 
safely assume that the quantum correction to the 
cosmological constant can be either negligible or have 
the form 
\beq
\de\rho_\La (observable) \,\, \sim \,\, 
H_0^2 M^2\,,
\label{CC-Quantum}
\eeq
where $M$ is the typical mass in the theory from where 
the corresponding quantum correction comes. Looking at 
the numerical part, it becomes clear that any physics 
below the GUT scale should be irrelevant for the quantum 
contribution to the vacuum energy.
Indeed, for $M=M_X \approx 10^{16}\,GeV$ the ratio 
between the hypothetical loop contribution and the value of
observable $\rho_\La$ or $\rho_c$ is about six orders of
magnitude. This is, in principle, a detectable effect
\cite{CCwave}, because it produces an essential difference 
in the density perturbations spectrum. However, already 
for the Standard Model we have 
$M=M_F\approx 10^2 - 10^{3}\,GeV$ and the difference grows 
up to the astonishing fifty orders of magnitude. For the 
vacuum effects of QCD the effect is even much weaker. 
The conclusion is that, contrary to our intuition, 
the physics below the GUT scale is likely irrelevant for 
the potential scale dependence of the vacuum energy. 
Indeed, this consideration may be applied 
only to the case when quantum corrections to the vacuum energy
do depend on the metric derivatives (or, better say, on Hubble
parameter). In principle, one can consider the case when the 
quantum corrections do not have such dependence 
\cite{cosm,bauer}. We shall give more details about the
possible running of vacuum energy density in section 7. 

\subsection{Calculational methods in curved spaces}

\qquad 
In this section we shall discuss the existing practical 
methods of quantum calculations in curved spaces and their 
relevance for establishing the general features of 
renormalization and structure of finite parts of quantum 
corrections. We shall avoid the detailed historical 
considerations which can be found in \cite{GMM,birdav,book}
and only discuss those aspects which are relevant for the 
consequent sections. We shall start by discussing the 
three methods (flat-space Feynman diagrams, local 
momentum representation and Schwinger-DeWitt method) 
which enable one to establish 
the general structure of renormalization in curved 
space-time. 

\vskip 2mm

$\bullet$ \
Feynman diagrams for the perturbations on flat background 
\ $h_{\mu\nu}=g_{\mu\nu}-\eta_{\mu\nu}$. Very important 
early works in this area have been done using this method 
\cite{UtDW,stze71}. In particular, these calculations 
have shown, for the first time, the necessity of higher 
derivative terms (\ref{HD}) for renormalizability and the 
general structure of finite quantum corrections, for both 
massive and massless cases. 

The contributions to the vacuum effective action
correspond to the diagrams with internal loops of matter 
fields and with 
external lines of the $h_{\mu\nu}$ field. One can
expand the action $S[\Phi,g]$ such that the propagators 
and vertices of all the fields (quantum and background) are 
the usual ones in the flat space-time. The internal lines 
of all the diagrams are only those of the matter fields, 
while external lines are both of matter and gravitational 
field $\,h_{\mu\nu}\,$. As a result,
any flat-space digram gives rise to the infinite set of  
diagrams, with increasing number of the background fields
tails. An example of such set is depicted in Figure 1. 

Due to the non-polynomial nature of gravity, every 
flat-space diagram is producing an infinite number of 
diagrams 
with external $h_{\mu\nu}$ tails, including an infinite 
number of divergent diagrams. In order to establish the 
structure of possible divergences, one has to impose
covariance which does not follow automatically from the 
calculations. Then the number of relevant diagrams 
becomes finite, making  the analysis of renormalization 
rather simple.  
The shortcomings of this method are the lack of explicit  
covariance, difficulties in practical calculations and 
interpreting the results. However, there is a serious 
benefit coming from the fact that, in the framework of 
this method, we deal with the usual flat-space QFT. 
According to the general theorems of the renormalization 
theory (see, e.g., \cite{vortyu,gomwein} and 
\cite{Weinberg-QFT}) the divergences in QFT can be removed 
by the local covariant counterterms. So, let us notice that 
the necessary counterterms in curved space are local ones. 

\vskip 2mm
\begin{figure}\qquad\qquad\qquad
\includegraphics[width=0.6\textwidth]{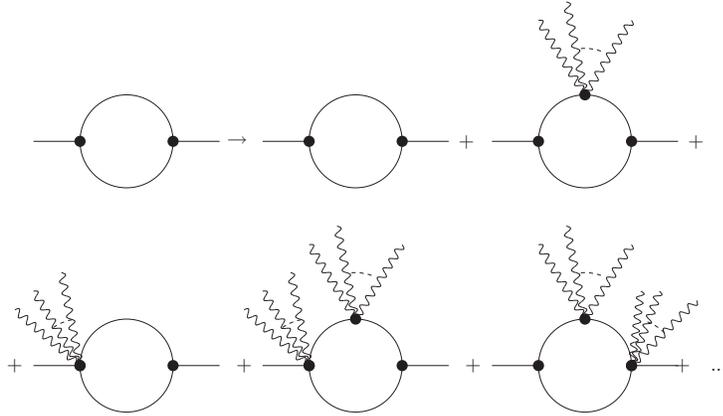}
\\
\caption{\sl 
 The straight lines correspond to the matter (in this case 
 scalar with $\,\lambda\varphi^3\,$ 
 interaction) field, and wavy lines to the external metric. 
 A single diagram in flat space-time generates an infinite 
 set of families of diagrams in curved space-time.
 The first of these generated diagrams is exactly the one in the
 flat space-time, and the rest have external gravity lines. }
\end{figure}
\vskip 3mm

$\bullet$ \ 
The simplest of the known covariant methods is based on 
the local momentum representation \cite{localMom}. In 
general, the use of momentum representation is not 
possible in curved spaces, however one can use normal 
coordinates (see, e.g. \cite{Petrov} for the introduction) 
and perform loop calculations in the tangent space 
corresponding to a given point $\,P\big(x^\mu_{(0)}\big)$. 
One can express all quantities of interest, such as free 
and interaction parts of the classical Lagrangian in form 
of the power series in $y^\mu=x^\mu-x^\mu_{(0)}$. Due to 
the special choice of coordinates, the coefficients of this 
series are components of curvature tensor and its covariant 
derivatives in the point $P$. The local momentum 
representation can be used for both propagator and vertices
and finally enables one to derive local quantities such 
as counterterms. For instance the first order expansion 
for the propagator of a massive non-minimal scalar field 
has the form 
\beq
G(k)\,=\,\frac{1}{k^2+m^2}
-\frac13\,\frac{(1 - 3\, \xi) R}{(k^2+m^2)^2}
+\frac23\,\frac{R^{\mu\nu}k_\mu k_\nu}{(k^2+m^2)^3} 
+{\cal O}(k^{-3})\,.
\label{scalar 7}
\eeq
Similar expansions can be constructed for spinor and 
vector fields and also for the interaction vertices. 
Taking the locality of the counterterms into account 
it is obvious that, after replacing such expansions
into Feynman diagrams, the structure of divergences 
will be as follows. The zero order terms will produce 
the divergences which are direct covariant generalizations 
of the ones in flat space-time. Furthermore, the quadratically 
divergent in flat space diagrams will produce logarithmic 
divergences (and therefore counterterms) linear in curvature 
tensor. Due to the dimensional reasons and gauge invariance, 
there are only two types of such terms. 
The first one is the Einstein-Hilbert type counterterm, 
with the coefficient proportional to the square of the 
mass of the quantum field. The second one is the non-minimal 
scalar-curvature interaction term which really shows up 
in the interacting scalar theory. Finally, in the next 
order in curvature there will be the fourth derivative 
counterterms. They can be constructed from the metric 
only, because matter fields are dimensional. The most 
important observation is that, if the original QFT in 
flat space is a renormalizable theory, starting from the 
third order in curvatures there will be only finite 
contributions. 

Let us notice that the locality of the counterterms 
follows from the previous 
\ $h_{\mu\nu}=g_{\mu\nu}-\eta_{\mu\nu}$ \ approach 
and from the universality of the vacuum EA. 
The information provided by the two methods which we 
briefly described above enables one to establish the 
general structure of renormalization in curved 
space-time. Another advantage is that these methods 
can be applied at higher loops. However, for the 
practical one-loop calculations we better apply 
another approach. 
\vskip 2mm

$\bullet$ \ \
Schwinger-DeWitt expansion \cite{BDW-65} is the most useful  
method  for calculating divergences and related quantities. 
The key point is the representation of the 
$\,\Tr \ln{\hat H}=\ln\Det{\hat H}\,$
via the proper time integral
\beq
\frac{i}{2}\,\Tr\ln {\hat H}
\,=\,-\,\frac{i}{2}\,\,\Tr\,
\int_{0}^{\infty}\,\frac{ds}{s}\,e^{ -is\,{\hat H} }\,,
\label{SDW}
\eeq
\beq
\mbox{where}
\qquad
e^{-is\,{\hat H}}={\hat U}_(x,x^\prime ;\,s)
\,=\,{\hat U}_0(x,x^\prime ;\,s)
\,\sum_{k=0}^{\infty}\,(is)^k\,{\hat a}_k(x,x^\prime)\,.
\label{heat}
\eeq
${\hat a}_k(x,x^\prime)$ \ are Schwinger-DeWitt 
coefficients, while the expression for 
${\hat U}_0(x,x^\prime ;\,s)$ has the form  
\beq
{\hat U}_0(x,x^\prime\,;s)\,=\,
\frac{1}{(4\pi i\,s)^{n/2}}\,\,{\cal D}^{1/2}(x,x^\prime)\,
\exp\left\{-\,\frac{\si(x,x^\prime)}{2i\,s}-im^2s \right\}\,.
\label{U0}
\eeq
Here $\,\si (x,x^\prime)\,$ is the geodesic distance between
$\,x\,$ and $\,x^\prime\,$ and \ 
$$
{\cal D}(x,x^\prime)
=\det \left[-\pa_\mu\pa_\nu \si(x,x^\prime)\right]
$$
is the Van Vleck-Morett determinant. Further details of the 
standard Schwinger-DeWitt technique can be found, e.g., in 
\cite{BDW-65}. A very powerful generalization, which enables 
one to derive 1-loop divergences for a variety of quantum 
theories on curved background and also for various models of 
quantum gravity, has been developed in \cite{bavi85} (see 
also references therein). A comprehensive review on the 
effective action method and in particular generalizations 
of the Schwinger-DeWitt technique has been recently given 
in \cite{Vilk08}.

\subsection{One-loop divergences and renormalization}

\qquad
In the framework of the Schwinger-DeWitt technique the UV 
limit corresponds  to the lower $s=0$ limit in the 
proper-time integral (\ref{SDW}). The regularizations of 
this integral can be performed in different ways. The most 
common is the special version of dimensional regularization 
\cite{bro-cass,bavi85}, but one can use also 
Zeldovich-Starobinsky approach \cite{stze71}, an 
equivalent adiabatic regularization \cite{adiabat}, 
point-splitting \cite{chris} or even 
covariant cut-off \cite{anom} regularizations. In the four 
dimensional case the ${\hat a}_0$-coefficient of (\ref{SDW}) 
corresponds to the quartic diveregence and the 
${\hat a}_1$-coefficient corresponds to the quadratic divergence.
It is well-known that the cancelation of these divergences 
may be related to the fine-tunings in imposing the 
renormalization conditions. This may be lead to difficulties 
in such cases as hierarchy problem in the Standard Model 
or the Cosmological Constant problem (which is, actually, 
also a hierarchy problem \cite{nova}) but we will not deal 
with these issues here. The reason is that, despite the 
fine tuning is unnatural, in the case of quadratic 
or quartic divergences it has no apparent concern for the 
form of quantum corrections which we are interested in.

The most important for us are logarithmic divergences 
which are proportional to the ``magic'' coefficient
$\,{\hat a}_2
= \Tr \lim\limits_{x^\prime \to x} {\hat a}_2(x,x^\prime)$ 
of (\ref{SDW}). 
The logarithmic divergences define such important notions as 
renormalization group $\,\beta$-functions and anomalies.
The form of $\,{\hat a}_2\,$ in the vacuum sector is
\beq
{\hat a}_2 = \int d^4 x\sqrt{g}\,\Big\{\,\be_\La + \be_E R 
+ \be_1C^2+\be_2E+\be_3{\Box}R+\be_4R^2
\,\Big\}\,.
\label{a2}
\eeq
where\footnote{We are introducing the double notations
\ $\be_{1,2,3}=(\om,b,c)/(4\pi)^2$. This will prove 
useful in the sections devoted to conformal anomaly and 
its applications.} 
\beq
(4\pi)^2\,\be_\La &=& \frac12\,N_2m_s^4 - 2N_fm_f^4\,\,,
\nonumber
\\
(4\pi)^2\,\be_E &=& N_sm_s^2\Big(\xi-\frac16\Big)\,
+\,\frac{N_f m^2_f}{3}\,,
\nonumber
\\
(4\pi)^2\,\be_4 &=& \frac{N_s}{2}\,\Big(\xi-\frac16\Big)^2\,,
\nonumber
\\
(4\pi)^2\,\be_1
 &=& \frac{1}{120}\,N_s + \frac{1}{20}\,N_f + 
\frac{1}{10}\,N_v \,=\, \om \,,
\nonumber
\\
(4\pi)^2\,\be_2 &=& -\,\frac{1}{360}\,N_s 
- \frac{11}{360}\,N_f - \frac{31}{180}\,N_v \,=\,b\,,
\nonumber
\\
(4\pi)^2\,\be_3 &=& \frac{1}{180}\,N_s
+ \frac{1}{30}\,N_f - \frac{1}{10}\,N_v \,=\,c\,.
\label{abc}
\eeq
Using these formulas, one can write the total expression for 
the divergent part of the one-loop effective action of vacuum 
for the theory involving $N_s$ real scalars, $N_f$ Dirac 
spinors and $N_v$ massless vectors 
\beq
{\bar \Ga}^{(1)}_{div}=-\frac{1}{n-4}\,
\int d^4x\sqrt{-g}\left\{\be_1 C^2
+ \be_2 E +\be_4R^2 + \be_3{\Box}R
+ \be_E R + \be_\La\right\}\,,
\label{tot}
\eeq
In (\ref{tot}) we used the notations of dimensional 
regularization, but of course the leading logarithmic 
divergence is regularization independent. 
\vskip 2mm

Let us make a few observations concerning the expression 
for the one-loop divergences (\ref{tot}). 

{\large\it i)} \ The divergences have exactly the 
form which follows from the general arguments presented 
above. All high derivative terms of dimension four, 
Einstein-Hilbert term and cosmological constant are 
necessary to have a renormalizable theory of matter 
on curved background. Let us remark that in case the 
interacting scalar is present, the non-minimal term 
$\xi R\ph^2$ is also necessary \cite{book} (see the 
references to the original papers therein). 

{\large\it ii)} \ According to the consideration 
presented above, at any loop order the divergences 
of the gauge theory in curved space-time are of the 
same form as the non-minimal classical action with 
the vacuum term or, in other words, of the same 
form as one-loop expressions. The only difference is 
the coefficients of the divergent terms, which start 
depending on the couplings at higher loops (see, e.g., 
\cite{hath82}). Therefore 
the theory formulated above is multiplicatively 
renormalizable in curved space-time. At any loop order 
the counterterms can be removed by renormalizing the 
full set of parameters of the theory, including couplings, 
masses, $\xi$ and vacuum parameters. There is no 
need to renormalize the external metric field. 

{\large\it iii)} \ The one-loop divergences are given 
by an algebraic sum of the contributions from a free
scalar, fermion and vector fields. The reason is that, 
as we have already mentioned above, each of the Feynman 
diagrams which we take into account consists of 
the single loop of a matter field with external tails 
of a gravity field. No matter vertices show up 
at this level. 

{\large\it iv)} \ In the massless case only the higher 
derivative counterterms emerge. This means the {\sl global} 
conformal symmetry holds in the one-loop divergences. As 
we shall see below, however, this symmetry is violated in 
the finite part of effective action.

{\large\it v)} \ The non-minimal parameter $\xi$ 
affects the one-loop divergences, but it can be only 
seen in the 
combination $\xi-1/6$. As we already know, the value 
\ $\xi=1/6$ \ corresponds, in the massless case, to the 
special version of the scalar theory which possesses the 
local conformal symmetry. 
Furthermore, the unique non-conformal counterterm, 
in the massless case, 
\ $\int \sqrt{-g}R^2$, \ has coefficient $(\xi-1/6)^2$. 
This means the one-loop divergences are conformal invariant 
if the original theory is invariant. More precise is to say
that the four-dimensional coefficient of the pole term 
satisfies the conformal Noether identity (\ref{Noether}). 

{\large\it vi)} \ All features described in the previous 
points can be explained in a systematic way. But there
are some amazing things in the expressions (\ref{abc}) 
which are still waiting to be explained. It is easy 
to see that the contributions of scalars, fermions 
and vectors to the $\be$-functions $\be_1$ and $\be_2$
show a universality of signs. The ones to $\be_1$ are 
all positive and the ones to $\be_2$ are all negative.  
As we shall see later on, the universality of the signs 
of $\be_2$ has great importance for the Starobinsky 
inflationary model. The mentioned sign rules have 
nothing to do with the Grassmann parity of the fields 
and remain an unexplained occurrence. It is even more 
mysterious that the higher derivative conformal fields 
(scalars and fermions) constructed so far 
\cite{rie,frts-SUGRA,cofe} always produce opposite 
signs (compared to the usual scalar, fermion and vector) 
of the contributions to both $\be$-functions. 

{\large\it vii)} \ Coming back to the heat kernel 
expansion (\ref{heat}) one can notice that the finite 
terms start from 
$\,\Tr \lim\limits_{x^\prime \to x} {\hat a}_k(x,x^\prime)
,\,\,\,k\geq 3\,$ 
and have the form of local covariant expressions with 
growing powers of metric derivatives. For instance, 
$a_3$ has the $R^3_{...}$-type and 
$R_{...}\Box R_{...}$-type terms. The general expressions 
are known for ${\hat a}_3$ and ${\hat a}_4$ 
\cite{Gilkey,Avramidi-90}.

\subsection{$\overline{\rm MS}$ scheme - based 
renormalization group}

\qquad
One can use the multiplicative renormalizability of the 
QFT in curved space-time to formulate the renormalization 
group equations for all quantum fields and parameters. 
As usual, the simplest version of the renormalization 
group can be constructed in the framework of the minimal
subtraction ($\overline{\rm MS}$) renormalization scheme 
\cite{nelspan82,buch84,Toms83}. The detailed exposition 
of the $\overline{\rm MS}$ scheme - based renormalization 
group method in curved space-time can be found in \cite{book}. 
Here we shall only give brief account of this method.

Consider the quantum theory of the fields $\Phi$ on the 
background of classical metric $g_{\mu\nu}$. The full 
set of parameters of the theory will be denoted by $P$.
For example, in case of SM or GUT - like QFT, $\Phi$ 
includes fermions, Yang-Mills fields and scalars and 
$P$ includes gauge, Yukawa and scalar self-interaction 
couplings, non-minimal parameters $\xi$ and the
parameters of the vacuum action (\ref{vacuum}), 
including Newton and cosmological constants.  
The dimensions of $\Phi$ and $P$ will be denoted 
as $d_\Phi$ and $d_P$, correspondingly. 

For the sake of definiteness we assume the dimensional 
regularization. The $\overline{\rm MS}$-scheme 
renormalization group equation for the effective action 
means that the last is independent on the dimensional 
renormalization parameter $\mu$
\beq
\mu\frac{d\Ga}{d\mu}\,=\,\left\{\,
\mu\frac{\pa}{\pa\mu}+\be_P(n)\,\frac{\pa}{\pa P}
+\int d^nx\ga_\Phi(n)\,\frac{\de}{\de \Phi (x)}
\,\right\}\,\Ga[g_{\al\be},\Phi,P,n,\mu]\,=\,0\,,
\label{n12}
\eeq
where we used the standard notations for the renormalization 
group $\be$ and $\ga$ - functions in $n$-dimensional 
space-time
\beq
\be_P(n)\,=\,\mu\frac{dP }{d\mu}\,,\qquad
\ga_\Phi(n)\,=\,\mu\frac{d\Phi}{d\mu}\,,
\label{n11}
\eeq
while the usual $\be$-functions correspond to the limit 
\ $n \to 4$. 
Let us notice that these formulas are essentially the 
same as in the flat space-time. The main difference is
the presence of $\,g_{\mu\nu}(x)$, playing the role of 
external parameter. As a result there are several
qualitatively new effective charges, such as $\xi$ and 
vacuum parameters. 

The eq. (\ref{n12}) is a formally universal renormalization 
group equation which can be used for different purposes, 
depending on the physical interpretation of $\mu$. For 
example, in order to consider the short distance limit, 
we can perform a global rescaling of all quantities, 
including metric and obtain another identity, which is 
independent on (\ref{n12})
\beq
\Ga[g_{\al\be},\Phi,P,n,\mu]\,=\,
\Ga[e^{-2\tau}g_{\al\be},\,
e^{d_\Phi \tau}\Phi,\,e^{d_P \tau}P,\,n,\,e^{\tau}\mu]\,.
\label{n14}
\eeq
After being considered together, (\ref{n12}) and (\ref{n14})
produce the general solution of the form 
\beq
\Ga[g_{\al\be}e^{-2\tau},\Phi,P,n,\mu]\,=\,
\Ga[g_{\al\be},\Phi(\tau),P(\tau),n,\mu]\,,
\label{n21}
\eeq
where the effective charges \ $P(\tau)$ and $\Phi(\tau)$ 
\ satisfy the renormalization group equations
\beq
\frac{d\Phi}{d\tau}\,=\,(\ga_\Phi-d_\Phi)\Phi\,,
\,\,\,\,\,\,\,\,\,\,\,\,\,\,\,
\frac{dP}{d\tau}\,=\,\be_P-Pd_P  \,.
\label{n22}
\eeq
The limit $\,\tau\to\infty\,$ corresponds to the short 
distances and, due to 
\beq
R^2_{\mu\nu\al\be}\to R^2_{\mu\nu\al\be} e^{4\tau}\,,
\,\,\,\,\,\,\,\,
R^2_{\al\be}\to R^2_{\al\be}e^{4\tau}\,,\,\,\,\,\,\,\,\,
R \to R e^{2\tau}\,,\,...
\label{n15}
\eeq
to the limit of greater curvatures. In this respect it 
is equivalent to the standard rescaling of momenta in the 
flat-space quantum field theory. On the other hand,
the application of (\ref{n21}) and (\ref{n22}) to 
particular situations needs a special attention. For 
example, let us consider the exponential inflation. 
The time-dependence of the metric 
\beq
g_{\al\be}\,\to\,g_{\al\be}\cdot e^{Ht}\,,\,\,\,\,\,\,
\,\,\,\,\,\,
H=const
\label{n23}
\eeq
looks similar to the rescaling (\ref{n21}). However, 
inflation does not fit the $\overline{\rm MS}$-based 
renormalization group, because the scalar curvature
$R=-12H^2$ does not behave according to (\ref{n15}). 
The origin of the difference is that the 
parameter of the {\it global} rescaling $\tau$ is a
constant, while the time $t$ in (\ref{n23}) is a 
coordinate and therefore the transformation (\ref{n23}) 
is a {\it local} one. 

In order to understand the physical significance of the 
renormalization group running in curved space one has 
to attribute some physical sense to the parameter $\mu$. 
At this point one has several option. Let us 
formulate the above question in a different way: which 
terms in the effective action can be parametrized by 
the $\overline{\rm MS}$-scheme renormalization group
in curved space? Remember that the standard 
interpretation of the $\overline{\rm MS}$-scheme 
renormalization group in flat space is related to the 
high energy limit in the momentum subtraction scheme
of renormalization. In this case, in order to get the 
terms in effective action which are behind the 
renormalization group, one has to replace 
$\,\ln\left(\mu^2/\mu^2_0\right)\,$ by the expression
$\,\ln(p^2/p^2_0)$, where $p^2$ is the square of the 
momentum. In the coordinate representation this means
we have to introduce the formfactor 
$\,\ln\left(\Box/m^2\right)$. 
For example, in case of QED, the corresponding term
looks like 
\beq
\be\,F^{\mu\nu}\,\ln \Big(\frac{\Box}{m^2}\Big)\,F_{\mu\nu}\,.
\label{formQED}
\eeq
This procedure can be indeed generalized to the case of 
a curved space. The expected term is, for instance,
\beq
\be_1\,C^{\mu\nu\al\be}\,\ln \Big(\frac{\Box}{m^2}
\Big)\,C_{\mu\nu\al\be}
\label{formWeyl}
\eeq
for the case of the Weyl term in the action of 
vacuum (\ref{vacuum}). Of course, the d'Alembertian 
operator in  (\ref{formWeyl}) is the covariant one. 
Later on, in section 6, we shall confirm the 
presence of the term (\ref{formWeyl}) by a direct 
calculation \cite{apco}. The $\overline{\rm MS}$-scheme 
based procedure described above can be successfully 
applied to derive the quantum corrections to the 
classical action of gravity and, e.g., scalar field 
\cite{buwolf,book} (see also \cite{Maroto} for an 
alternative consideration). At the same time, this approach 
meets obvious difficulties when applied to the running 
of the Newton or cosmological constants 
\cite{apco,barv03,Mazz}.
In these cases the formfactor can not be simply inserted 
into the action, because the d'Alembertian operator acting
on the cosmological constant gives zero. Similarly, in 
the Einstein-Hilbert term, the formfactor can only produce 
the total derivatives, or superficial terms, which do not 
affect the equations of motion for gravity. We shall 
come back to these arguments in section 7. 

Can we state that the identification of $\mu^2$ with 
$\Box$ is a universal tool making the 
$\overline{\rm MS}$-scheme results physically relevant? 
Unfortunately, the problem is not really solved by the 
analogy with QED, because in gravity, according to the 
relations (\ref{n15}) one can identify $\mu$ with 
other metric-dependent quantities. 

For example, the scalar curvature has the same global 
scaling law as the d'Alembertian operator $\Box$. In this 
way we can construct the curvature-controlled version of 
renormalization group. It is interesting that the 
corresponding expressions in the effective action 
can be also obtained by special resummation of the 
Schwinger-deWitt series
\cite{Parker-resum85,Parker-resum}. From the general 
perspective the identification of $\mu^2$ with the 
scalar curvature is as legitimate as its identification 
with $\Box$ and moreover it has the following two 
advantages: the possibility to write down the quantum
correction to the cosmological and Newton constants and 
the closeness to the most natural identification of 
$\mu$ with the Hubble parameter in the cosmological 
setting \cite{scale-setting,Gruni}. One has to 
remember, however, that anyone of these identifications 
is no more than a particular model for an unknown 
complete effective action. An obvious manifestation 
of this feature is a vast ambiguity which one 
meets in the resummed and truncated expressions used 
in \cite{Parker-resum}.
It is important to remember that the renormalization 
group is a method to parametrize the scaling dependence 
of the effective action. 
In case of gravity it is especially difficult to introduce 
a physically consistent and universal renormalization 
group, in particular because definition of the energy
of the gravitational field is a nontrivial problem
(see, e.g., \cite{DeTe} and references therein). 

The solution of the renormalization group equations 
(\ref{n22}) has been explored for different QFT models 
in curved space \cite{book} (see further references 
therein). The equations for 
couplings and masses of the quantum matter fields are 
exactly the same as in flat space-time \cite{ChengLie}. 
The qualitatively new elements are the equations for 
the nonminimal parameters $\xi$ and for the parameters 
of the vacuum action \ $\La,\,G,\,a_{1,2,3,4}$.
The one-loop $\be_\xi$ is always proportional to 
the difference $\xi-1/6$, such that the conformal 
value \ $\xi=1/6$ \ is a universal fixed point of the 
renormalization group trajectory in the theory with 
scalar fields. The coefficient of proportionality 
depends on the model. For some gauge models it 
leads to the UV asymptotic conformal invariance 
\cite{buod85} and for other models the conformal value 
is the IR fixed point. The general conditions for the 
asymptotic behavior of $\xi$ have been found in 
\cite{yagun}. Indeed at higher loops the $\be_\xi$ 
is not proportional to $\xi-1/6$ factor anymore 
\cite{hath82}. The related ambiguities lead to 
the quantum inconsistency of the conformal invariant 
theory \cite{anom} beyond the one loop approximation. 

Finally, there is one subtlety in the physical 
application of the renormalization group equations for 
the cosmological and Newton constants (\ref{abc})
\beq
(4\pi)^2\,\mu\,\frac{d}{d\mu}\left(\frac{\La}{8\pi G}\right)
\,=\, \be_\La - \frac{\La}{2\pi G}
\,,\qquad
\mu\,\frac{d}{d\mu}\left(\frac{1}{16\pi G}\right)
\,=\, \be_E - \frac{1}{8\pi G}\,.
\label{RG-CCG}
\eeq
Do we need the second terms in the {\it r.h.s.} of these
two equations?  In order to address this question one has 
to remember that $\La$ and $G$ gain physical sense when 
they are inserted into the Einstein equations 
\beq
R_{\mu\nu}-\frac12\,Rg_{\mu\nu}=8\pi GT_{\mu\nu}
+ \La g_{\mu\nu}\,.
\label{Einstein}
\eeq
The second terms of (\ref{RG-CCG}) reflect the classical 
dimension of $\La$ and $G$, but other components of 
(\ref{Einstein}) also have their own dimensions. As a 
result the classical scaling terms do cancel and, for 
the applications, one has to use only anomalous 
dimensions, that are the $\be$-functions terms. 

\section{Conformal anomaly and anomaly-induced EA}

\qquad
In general, there is no way to calculate the vacuum effective action
completely. In order to understand why this is so, let 
us remember that already in the Schwinger-DeWitt expansion 
we observe an infinite series in curvature tensor and 
its derivatives. On the top of that, the renormalization 
group and analogy with QED indicates, as we have 
discussed in the previous section, the presence of 
an infinite amount of non-local insertions. So, 
the task of deriving the full effective action does 
not look realistic. It is remarkable that there is a
class of the four-dimensional theories for which we are 
able to derive exactly the non-local terms in 
\ ${\bar \Ga}^{(1)}$. The word ``exactly'' here
means we can do it at the one-loop level and only 
on the very special backgrounds. Despite of
these restrictions, the situation deserves this wording, 
as a way to emphasize an enormous difference 
with the situation typical for other kinds of theories. 

In this section we consider the anomalous violation
of local conformal symmetry in the case of quantum matter
on classical curved background. The conformal anomaly 
enables one to derive the anomaly-induced effective 
action. In the consequent section we shall discuss the 
applications of anomaly and of the induced effective 
action. 

\subsection{Derivation of conformal anomaly}

\qquad
Quantum anomaly is a typical phenomenon in a situation 
where the original theory has more than one symmetry. 
The origin of anomalies is the renormalization procedure
\cite{Peskin}. The anomaly shows up if there is no 
regularization which preserves all symmetries at the
quantum level. After the divergences 
are subtracted, in the finite part of the effective 
action some of the symmetries are getting broken. In 
our case the theory has general covariance and local 
conformal symmetry, and the last is broken by quantum 
corrections. 

Consider quantization of free massless conformal 
invariant matter fields denoted by $\Phi$, on classical 
gravitational background. As before, we assume that the 
set of quantized matter fields $\Phi$ includes
\ $N_s$ scalars (all with $\xi=1/6$), $\,N_f\,$ fermions 
and $\,N_v$ \ vectors. We denote \ $k_\Phi$ \ the 
conformal weight of the field.

At the one-loop level
it is sufficient to consider the simplified vacuum action
\beq
S_{conf.\,\,vac}\,=\,\int d^4x \sqrt{-g}
\left\{a_1C^2+a_2E+a_3{\Box}R \right\}\,.
\label{S_vac-conf}
\eeq
Let us emphasize that it is not wrong to supplement the
last expression by the Einstein-Hilbert action, cosmological 
constant or the \ $\int\sqrt{-g}R^2$-term. The action 
(\ref{S_vac-conf}) can be seen as a part of classical 
action which is a subject of an infinite renormalization 
at the one-loop level. Beyond the one-loop approximation  
the \ $\int\sqrt{-g}R^2$ \ term is also necessary, 
because the conformal theory becomes inconsistent 
\cite{anom}. 

The Noether identity for the local conformal symmetry has
the form (\ref{Noether}) and can be interpreted as
$\,T^\mu_\mu=0\,$ on shell (see section 2).
At quantum level $\,S_{vac}(g_{\mu\nu})\,$ has to be
replaced by the effective action of vacuum
$\,\Ga_{vac}(g_{\mu\nu})$. As we already know, its 
divergent part is
\beq
\Ga_{div}\,=\,\frac{1}{\vp}\,\int d^4x\sqrt{g}
\left\{\be_1C^2+\be_2E+\be_3{\Box}R \right\}\,,
\label{diverg}
\eeq
where \ $\vp=(n-4)^{-1}$ \ in dimensional regularization.
In the case of global conformal symmetry, the renormalization
group method or \ $\ze$-regularization tell us
\cite{haw-z,buch84,book}
\beq
<T^\mu_\mu>\,=\,\be_1C^2+\be_2E+a^\prime{\Box}R \,,
\label{T}
\eeq
where $a^\prime=\be_3$. In the case of local conformal
invariance there is an ambiguity in the parameter \ $a^\prime$ 
\ \cite{birdav,duff94,AGS}. This issue has been explained 
recently in \cite{anom} and reviewed in \cite{Conf-Proc}, 
so we will not discuss it in full details here. Qualitatively 
the result is that the ambiguity is always equivalent to 
the freedom to add the \ $\int\sqrt{-g}R^2$-term to the 
classical action. 

One can derive the conformal anomaly in different ways, 
mainly due to the choice of regularization schemes 
\cite{duff77,chris,birdav,anom}. We shall follow 
\cite{duff77} and \cite{anom,Conf-Proc}, using dimensional regularization. We are interested in the vacuum effects 
and therefore, at the one-loop level, can restrict 
consideration by the free fields case. The expression 
for divergences is (\ref{diverg}) with the $\be$-functions 
defined in (\ref{abc}). 
The renormalized one-loop effective action has the form
\beq
\Ga_R = S + {\bar \Ga} + \De S\,,
\label{total}
\eeq
where \ ${\bar \Ga}={\bar \Ga}_{div}+{\bar \Ga}_{fin}$ \ 
is the naive quantum correction to the classical action and
$\De S$ is an infinite local counterterm which is called to 
cancel the divergent part of ${\bar \Ga}$ (\ref{diverg}). 
Indeed $\De S$ is the only
source of the noninvariance of the effective action, since
naive (but divergent) contributions of quantum matter fields
are conformal. The anomalous trace is therefore equal to
\beq
T = <T_\mu^\mu> = - \frac{2}{\sqrt{-g}}\,g_{\mu\nu}
\left.\frac{\de \,\Ga_R}{\de \,g_{\mu\nu}} \right|_{n=4}
= - \frac{2}{\sqrt{-g}}\,g_{\mu\nu}
\left.\frac{\de\,\De S}{\de\, g_{\mu\nu}} \right|_{n=4}\,.
\label{trace}
\eeq
The calculation of this expression can be done, in a most
simple way, as follows.
Let us change the parametrization of the metric to
\beq
{g}_{\mu\nu} = {\bar g}_{\mu\nu}\cdot e^{2\si}\,,\qquad
\si=\si(x)\,,
\label{conf}
\eeq
where $\,{\bar g}_{\mu\nu}\,$ is the fiducial metric with
fixed determinant. There is a useful relation
\beq
 - \frac{2}{\sqrt{-g}}\,g_{\mu\nu}
\frac{\de\,A[g_{\mu\nu}]}{\de\, g_{\mu\nu}}
= - \frac{1}{\sqrt{-{\bar g}}}\,e^{- 4\si}
\left.\frac{\de\,A[{\bar g}_{\mu\nu}\,e^{2\si}]}{\de \si}
\,\right|_{{\bar g_{\mu\nu}}\rightarrow g_{\mu\nu},
\si\rightarrow 0}\,.
\label{deriv}
\eeq   
At that point we need transformation laws for the
structures presented in (\ref{diverg}). They can be found,
for instance, in \cite{Stud}, so we will not reproduce
these formulas here. For instance, for the square of the
Weyl tensor we have
\beq
\int d^nx \sqrt{-g}\,C^2(n)\,=\,
\int d^nx \sqrt{- {\bar g}}\,e^{(n-4)\si}\,{\bar C}^2(n)\,.
\label{c-trans}
\eeq
All other expressions of our interest (\ref{diverg}) have 
the same factor $\,e^{(n-4)\si}\,$ and, on the top of that, 
some extra terms with derivatives of $\,\si(x)$. For all 
terms which are not total derivatives, these terms are 
irrelevant due to the procedure (\ref{deriv}).

In the simplest case of global conformal factor
$\,\si=\la=const\,$ we immediately arrive at the expression
(\ref{T}) with $a^\prime=\be_3$. However in the local
case \ $\si=\si(x)$ \ the situation is more complicated.
It is worth mentioning that the left hand side in
(\ref{deriv}) gives zero when applied to the integral
of the total derivative term \ $\int\sqrt{-g}\Box R$.
On the other hand, the  value of  $\,a^\prime\,$ can be 
modified by adding a finite term 
\beq 
S_{+}\,=\,\al\,\int d^4x\sqrt{-g}\,R^2
\label{R^2}
\eeq
to the classical action (\ref{vacuum}), due to the identity
\beq
 - \frac{2}{\sqrt{-g}}\,g_{\mu\nu}
\frac{\delta }{\delta g_{\mu\nu}}\,\int d^4x\sqrt{-g}\,R^2
\,=\, 12\,{\square} R\,.
\label{identity}
\eeq
The last formula can be derived either directly or through 
the eq. (\ref{deriv}). The same effect can be achieved by 
the term $\int\sqrt{-g}R_{\mu\nu}^2$, 
\beq
 - \frac{2}{\sqrt{-g}}\,g_{\mu\nu}
\frac{\delta }{\delta g_{\mu\nu}}\,\int d^4x\sqrt{-g}\,
R_{\mu\nu}\,R^{\mu\nu}\,=\, 4\,{\square} R
\label{identityRicci}
\eeq
and also by the term $\,\int\sqrt{-g}R_{\mu\nu\al\be}^2$.
\ For the sake of simplicity, below we shall discuss only 
the term $\int\sqrt{-g}R^2$.

One may think that adding the classical non-conformal term 
(\ref{R^2}) has nothing to do with the quantum corrections.
However, consider in more details how to apply the procedure
(\ref{trace}) to the counterterm of the 
$\int\sqrt{-g}C^2$-type. The point is that the Weyl tensor 
depends on the dimension $d$, in particular its square is
\beq
C^2(d)\,=\,C_{\al\be\mu\nu}(d)\,C^{\al\be\mu\nu}(d)
\,=\,R^2_{\mu\nu\al\be}
-\frac{4}{d-2}\,R^2_{\mu\nu}+\frac{2}{(d-1)(d-2)}\,R^2\,.
\label{A14}
\eeq
When defining the corresponding counterterm
\beq
\De S_C\,=\,\frac{\be_1}{n-4}\,\int\sqrt{-g}C^2(d)\,,
\label{DeC}
\eeq
one can choose the Weyl tensor with $\,d=n+\gamma (n-4)$,
where $\,\ga\,$ is an arbitrary number. For any such $d$ the 
counterterm is local, it cancels the divergent part of 
effective action and the renormalization is multiplicative 
in the $R^2_{\mu\nu\al\be}$, $\,R^2_{\mu\nu}\,$ and $\,R^2$
basis. However, the anomalous $\,\Box R$ term depends on 
the choice of $\ga$. \ Hence at this point we meet an 
arbitrariness. The particular choice $\,\ga=-1\,$ has been
done in \cite{duff77}. In this case we arrive at
\beq
T(C^2)
\,=\,\frac{2}{\sqrt{-g}}\,g_{\mu\nu}\,\frac{\de}{\de g_{\mu\nu}}
\,\frac{\mu^{n-4}}{n-4}\,\int d^nx\sqrt{-g}\,\be_1\,C^2(4)
\Big|_{n\to 4} = C^2-\frac23\,{\Box}R\,.
\label{d=4}
\eeq
For scalars and spinors the result is identical for the 
global and local conformal symmetries violation. In case 
of vectors there is no such equivalence and, moreover, 
this is only one particular case of the possible 
counterterms. 
Finally, it is easy to see that the difference between 
the counterterms (\ref{DeC}) with different $d$ is equivalent 
to the {\it finite} $\int\sqrt{-g}R^2$-term. Qualitatively 
similar ambiguity takes place in the covariant Pauli-Villars 
regularization \cite{anom}. Finally, the anomaly is given 
by (\ref{T}), but there is an ambiguity in the 
coefficient $\,a^\prime$. If most of regularization 
schemes $\,a^\prime = \be_3$ but, in general, fixing 
this coefficient requires a special renormalization 
condition \cite{AGS,Conf-Proc}.

\subsection{Anomaly-induced action}

\qquad 
One can use conformal anomaly to construct the equation for
the finite part of the 1-loop correction to the effective
action (the notations are according to (\ref{abc}))
\beq
\frac{2}{\sqrt{-g}}\,g_{\mu\nu}
\frac{\de\, {\bar \Ga}_{ind}}{\de g_{\mu\nu}}
\,=\,\om C^2 + bE + c{\Box} R\,.
\label{mainequation}
\eeq
The solution of this equation is straightforward \cite{rie}
(see also generalizations for the theory with torsion
\cite{buodsh} and with a scalar field \cite{Shocom}).
Here we shall follow a bit more detailed exposition 
given in \cite{Conf-Proc}.
The simplest possibility is to parametrize metric as in
(\ref{c-trans}), separating the conformal factor \ $\sigma(x)$
\ and rewrite the eq. (\ref{mainequation}) using (\ref{deriv}).
The solution for the effective action is
\beq
{\bar \Ga} &=& S_c[{\bar g}_{\mu\nu}] \,+\,
\int d^4 x\sqrt{-{\bar g}}\,\{
\om \si {\bar C}^2 + b\si ({\bar E}-\frac23 {\bar {\Box}}
{\bar R}) + 2b\si{\bar \De}_4\si 
\nonumber
\\
&-& \frac{1}{12}\,(c+\frac23 b)[{\bar R} - 6({\bar \na}\si)^2 
- ({\bar \Box} \si)]^2)\}
\label{quantum}
\eeq
where $S_c[{\bar g}_{\mu\nu}]=S_c[g_{\mu\nu}]$ is an unknown
functional of the metric, which serves as an integration
constant for the eq. (\ref{mainequation}).
The merit of the solution (\ref{quantum}) is its simplicity,
but it is not covariant or, in other words, it is not 
expressed in terms of original metric $\,g_{\mu\nu}$. It 
proves useful to obtain the covariant solution \cite{rie,a}.
Let us establish the following relations \cite{rie}
(see also \cite{Stud} for details):
\beq
\sqrt{-g}C^2 = \sqrt{-{\bar g}}{\bar C}^2 \,,\,\,\,\,\,\,\,\,\,
 \sqrt{-{\bar g}}\,{\bar \De}_4 = \sqrt{- {g}}\,{\De}_4\,,
\label{weyly}
\eeq
\beq
 \sqrt{-g}(E - \frac23{\Box}R) = \sqrt{-{\bar g}}({\bar E}
- \frac23{\bar {\Box}}{\bar R} + 4{\bar {\De}}_4\si )\,,
\label{trans Gauss}
\eeq
where $\,\De_4\,$ is a fourth derivative conformally 
covariant operator acting on dimensionless scalar 
\beq
\De_4 \,=\, \Box^2 + 2R^{\mu\nu}\na_\mu\na_\nu
- \frac23\,R\Box +\frac13\,R_{;\mu}\,\na^\mu
\label{rei op}
\eeq
and also introduce the Green function for this operator
\ $\De_{4,x}\,G(x,y)=\de(x,y)$.
Using these formulas and (\ref{deriv}) we find, for any
$A = A({\bar g}_{\mu\nu},\si)$, the relation
\beq
\frac{\de}{\de \si (y)}\,\int d^4 x \sqrt{-g (x)}\,A\,
\left.(E
- \frac23{\Box}R)\right|_{g_{\mu\nu} = {\bar g}_{\mu\nu}}
 = 4\sqrt{-{\bar g}}{\bar {\De}}_4 \,A = 4\sqrt{- g}{\De}_4 \,A\,.
\label{GB}
\eeq
Using this relation it is easy to find the term in the effective 
action, which is responsible for $T_\om  = - \om  C^2$,
\beq
\Gamma_\om  = \frac{\om}{4}\,\int d^4 x \sqrt{-g (x)}\,
\int d^4 y \sqrt{-g (y)}\, C^2(x)
\,G(x,y)\,(E - \frac23{\Box}R)_y\,.
\label{a-term}
\eeq
Similarly one can find the term which
produces \ $T_b = b\,(E - \frac23{\Box}R)$
\beq
\Gamma_b = \frac{b}{8}\,\int d^4x \sqrt{-g (x)}\,
\int d^4 y \sqrt{-g (y)}\,
(E - \frac23{\Box}R)_x\,G(x,y)\,(E - \frac23{\Box}R)_y\,.
\label{b-term}
\eeq
The third constituent of the induced action follows from 
eq. (\ref{identity})
\beq
\Ga_{c} = - \frac{3c+2b}{36}
\,\int d^4 x \sqrt{-g (x)}\,R^2(x) \,.
\label{c-term}
\eeq
The covariant solution of eq. (\ref{mainequation})
is a sum of the expressions (\ref{a-term}), (\ref{b-term}) 
and (\ref{c-term}).

The anomaly-induced action presented in a local form using 
auxiliary scalar fields. After
using the classical equations of motion for these fields
and replacing them back to the action we come to the
original non-local expressions. 
In order to construct the local covariant representation,
the action should be presented in a symmetric form
\beq
\Gamma_{\om,b}
&=& \int d^4 x \sqrt{-g (x)} \int d^4 y \sqrt{-g (y)}
(E - \frac23{\Box}R)_x G(x,y)\left[\frac{\om}{4}C^2
- \frac{b}{8}(E - \frac23{\Box}R)\right]_y
\nonumber
\\
&=& -\frac{b}{8}\int d^4 x d^4 y \sqrt{g (y)g (x)}
\left[(E - \frac23{\Box}R) -
\frac{\om}{b}C^2\,\right]_x
G(x,y)\left[(E - \frac23{\Box}R) - \frac{\om}{b}C^2\right]_y
\nonumber
\\
&-& \frac12\,\int d^4 x d^4 y \sqrt{g (y)g (x)}\,
\left(\,\frac{\om}{2\sqrt{-b}}\,C^2\,\right)_x\,G(x,y)\,
\left(\,\frac{\om}{2\sqrt{-b}}\,C^2\,\right)_y\,.
\label{nonloc}
\eeq
The last two terms are appropriate objects for rewriting them
using the auxiliary fields. In this way we arrive at the following
final expression for the anomaly generated effective action of
gravity.
\beq
{\bar \Ga} &=& S_c[g_{\mu\nu}]
- \frac{3c+2b}{36}\,\int d^4 x \sqrt{-g (x)}\,R^2(x)
\label{finaction}
\\
&+& 
\int d^4 x \sqrt{-g (x)}\,\Big\{
\frac12 \,\ph\De_4\ph - \frac12 \,\psi\De_ 4\psi
\,+\, \ph\,\Big[k_1\,C^2
\,+\, k_2\,\big(E -\frac23\,{\Box}R\big)\Big]
+ l_1\,\psi\,C^2 \,\Big\}\,,
\nonumber
\eeq
where 
\beq
k_1 = - l_1 = - \frac{\om}{2\sqrt{-b}}
\qquad \mbox{and} \qquad
k_2 = \frac{\sqrt{-b}}{2}\,.
\label{kl}
\eeq
\vskip 1mm

Some important remarks are in order.

1) The local covariant form (\ref{finaction}) is dynamically
equivalent to the non-local one (\ref{nonloc}), (\ref{c-term}). 
The complete definition of the Cauchy problem in the theory 
with the non-local action requires defining the boundary 
conditions for the Green functions \ $G(x,y)$, independently
in the two terms \ (\ref{a-term}) \ and \ (\ref{b-term}). The
same can be achieved, in the local version, by imposing the
boundary conditions for the two auxiliary fields $\ph$ and 
$\psi$.

2) The local form of effective action with two auxiliary 
scalars (\ref{finaction}) has been introduced in the paper 
\cite{a}. Qualitatively similar manner of introducing second 
scalar has been suggested later on in \cite{MaMo}.
The kinetic term for the auxiliary field $\ph$ is positive
while for $\si$ it was negative. For $\psi$ the kinetic term
has negative sign. The wrong sign does not lead to problems
here, because both fields are auxiliary and do not propagate
independently.

3) We introduced the new structure
\ $\int C^2_x G(x,y) C^2_y$ \ into the action, despite it
was not directly produced by anomaly. This term is indeed
conformal invariant and therefore its emergence may be viewed
as a simple redefinition of the conformal invariant functional
$S_c[g_{\mu\nu}]$. On the other hand, writing the non-conformal
terms in the symmetric form (\ref{nonloc}), we have modified
the four point function in a very essential way. Therefore,
introducing the mentioned conformal term we have just restored
the basic structure of the terms generated by anomaly. For this
reason, the second auxiliary scalar \cite{a}
represents a natural element of writing the induced action 
in a local form.

\subsection{Light massive fields}

\qquad
Before we start to discuss the applications of conformal 
anomaly and of the anomaly-induced action, let us show 
how the same approach may be useful for exploring the 
effective action of light massive fields. In section 6 
we shall also consider the case of heavy massive fields. 
In order to understand why these two cases must be treated 
separately, let us come back and look at the Schwinger-DeWitt 
expansion (\ref{heat}). It is easy to see that it 
corresponds to the growing powers in metric derivatives. 
Due to the covariance, this is actually an expansion in 
the positive powers in curvatures and their covariant 
derivatives. The dimension is compensating by the 
negative powers of the mass $m$ of the quantum field. 
Indeed, such expansion is going to be efficient is
the curvature invariants are much smaller than $m^2$.  
In other 
words this is an approximation for the ``heavy'' fields
case. However, there are physical situations where we 
need another approximations, for instance for the case 
of light fields, where  curvature invariants satisfy
$\left|R_{....}\right| \ll m^2$, or for an intermediate 
regime, where these two quantities are of the same order 
of magnitude. It is remarkable that we have no regular 
approximations for these two cases and, in the last 
case, there is no available method at all. However, for 
the light fields, there is one method, suggested in 
\cite{Shocom} and slightly generalized in \cite{asta}. 

Let us consider the theory where the conformal invariance 
of scalar and fermion actions is violated only by the 
masses of these fields and by the Einstein-Hilbert 
action. The idea is to treat the masses as 
perturbations with respect to the massless theory. 
Originally, there is no local conformal symmetry and hence
one can not use the conformal anomaly to derive quantum 
corrections. But, this can be changed if we apply the 
conformal parametrization of the massive theory. The 
transformation is similar to the one used long ago 
in the cosmon model \cite{cosmon} and in GR \cite{deser70}. 
The calculations perform in the following three steps: 

1) Conformization of the theory; 

2) Derivation of anomaly and 
anomaly-induced effective action of the new theory; 

3) Coming back to the original parametrization. 

Let us replace the dimensional parameters in both matter 
and gravitational sectors by the powers of a new auxiliary 
scalar field $\,\chi$ according to  
\beq
m_s^2 \to \frac{m_s^2}{M^2}\,\chi^2
\quad\mbox{(scalar mass)}\,,
\qquad
m_f \to \frac{m_f}{M}\,\chi 
\quad\mbox{(fermion mass)}\,,
\nonumber
\\
\frac{1}{16\pi G}\,R \,\to\, \frac{M_P^2}{16\pi M^2}\,
\left[\,R\chi^2 + 6\,(\pa \chi)^2\,\right]
\,,\,\,\,\,\,\,\,\,\,\,\,\,\,\,\,\,\,\,\,\,\,\,\,\,\,\, 
\La \to \frac{\La}{M^2}\,\chi^2 \,.\quad\quad
\label{replace}
\eeq
Here $\,M\,$ is a new dimensional parameter. 
In order to provide the local conformal invariance we 
postulate that the auxiliary field $\chi$ transforms as
$\,\chi \to \chi\,e^{-\sigma}\,,\,\,\si=\si(x)$. Under 
the procedure (\ref{replace}), the massive  terms for the 
matter fields are replaced by Yukawa and scalar 
interactions involving the new auxiliary 
scalar $\,\chi$. The new action is conformal invariant
in both matter and gravitational sectors. The 
classical Noether identity for the new vacuum action 
has the form
\beq 
{\cal T}\, = \,\Big(-\,\frac{2}{\sqrt{-g}} g_{\mu\nu}
\,\frac{\de }{\de g_{\mu\nu}}
+  \frac{1}{\sqrt{-g}}\,\chi\,
\frac{\de }{\de \chi}\,\Big)
\,S_{vac}[g_{\mu\nu},\,\chi]\,= \,0\,.
\label{vacu}
\eeq

The conformal invariance is violated by a conformal 
anomaly and one can derive the effective action of the 
background fields $g_{\mu\nu}, \chi$ using the
anomaly-induced effective action scheme.
The divergences of the theory in the conformal 
representation have the form 
\beq
\Ga^{(1)}_{div} = - \frac{\mu^{n-4}}{(n-4)}\,
\int d^nx\sqrt{-g}\,\Big\{ wC^2 + bE + c{\nabla^2}R 
+\,\frac{f}{M^2}\,[R\chi^2 + 6(\pa \chi)^2]
\,\,+\,\,\frac{g}{M^4}\,\chi^4\Big\}\,,
\label{div}
\eeq
where the coefficients $w,b$ and $c$ are given by Eq.
(\ref{abc}) and 
\beq
f\,=\,\frac{1}{3(4\pi)^2}\,\sum_{f}\,{N_f\,m_f^2}
\,, \qquad
g\,=\,\frac{1}{2(4\pi)^2}\,\sum_{s} \,{N_s\,m_s^4}
-\frac{2}{(4\pi)^2}\sum_{f}\,{N_f\,m_f^4}\,.
\label{g}
\eeq
Here the sums are taken 
over massive fermion $f$ and scalar $s$ fields with 
the masses $m_f$ and $m_s$ correspondingly,  
$N_f$ and $N_s$ being multiplicities. 

The conformal anomaly is proportional to the integrand of 
(\ref{div})
\beq
<{\cal T}> \,=\, -\, \Big\{\,\,
wC^2\,+\,bE\,+\,c{\nabla^2}R \,+\,\frac{f}{M^2}
\,[R\chi^2+6(\pa \chi)^2]
\,+\,\frac{g}{M^4}\,\chi^4\,\,\Big\}\,.
\label{trace anomaly}
\eeq
In the standard way, described in section 5.1, one can 
derive the anomaly-induced effective action of the background 
fields $g_{\mu\nu}$ and $\chi$ 
\beq
\Ga_{ind} \,=\, S_c[g_{\mu\nu}, \chi]
\,+\,\int d^4 x\sqrt{-{\bar g}} \,\{w{\bar C}^2\sigma
+ b({\bar E} -\frac23 {\bar \nabla}^2 {\bar R})\sigma
+ 2 b\,\sigma{\bar \Delta}\sigma + 
\nonumber
\\
+ \frac{f}{M^2}\,[\,{\bar R}{\bar \chi}^2 + 6\,(\partial {\bar
\chi})^2]\sigma\, + \, \frac{g}{M^4}\,{\bar \chi}^4\sigma\,\} 
\,\,-\,\, \frac{3c+2b}{36}\,\int d^4x\sqrt{-g}\,R^2\,.
\label{quantum M}
\eeq
The last step is to fix the conformal unitary gauge
$\,\chi = {\bar \chi}\,e^{-\si}=M$, such that the 
classical Einstein-Hilbert and cosmological terms acquire 
their standard form
\beq
\Gamma 
& = & S_{HD}+ {\bar \Ga}_{ind}[{\bar g}_{\mu\nu},\,\si]
\label{quantum for massive}
\\
&-& \int d^4 x\sqrt{-{\bar g}} 
e^{2\si}[{\bar R}+6({\bar \na}\si)^2]
\cdot\Big(
\frac{1}{16\pi G} - f\cdot\si\Big)
- \int d^4 x\sqrt{-{\bar g}}e^{4\si}\cdot
\Big(\frac{\La}{8\pi G}-g\cdot\si\Big) \,,
\nonumber
\eeq
where $\,{\bar \Ga}_{ind}[{\bar g}_{\mu\nu},\,\si]\,$
has been defined in (\ref{quantum}). 

Let us discuss the level of reliability behind the 
calculation of the effective action 
(\ref{quantum for massive}). First of all, it is easy 
to see that this is not an exact result, even for the 
FLRW-like metric. The reason is that the integration 
constant $S_c[g_{\mu\nu}, \chi]$ in (\ref{quantum M}) 
depends on the intermediate scalar $\chi$ and, after 
fixing \ $\chi\to M$, \ $S_c[g_{\mu\nu},M]$ is not 
conformal invariant functional anymore. The second 
observation is that the higher-derivative part of the 
Eq. (\ref{quantum for massive}) is identical to that for 
the massless fields. Let us remember that the one-loop 
effective action is given by \ $(i/2)\ln\Det {\hat H}$
\ of some operator, depending on the background metric 
\cite{dewitt-75}. The \ $\Det{\hat H}$ \ can be seen 
as a product of the eigenvalues of the operator 
\ ${\hat H}$, \ depending on the corresponding 
eigenfunctions (see, e.g., \cite{elizalde}). When we 
are adding the masses of the fields, there are two 
distinct effects: modified eigenvalues due to the 
masses and also modified eigenfunctions. It is 
obvious that the calculation presented above takes 
into account only the first effect, that is why the 
higher derivative sector does not change. Perhaps 
the second effect manifests itself in the 
$S_c[g_{\mu\nu},M]$-term, but this part is out of 
our control. 

In order to understand better the approximation which 
is assumed in our calculus, let us notice a strong link 
between the anomaly-induced effective action 
Eq.(\ref{quantum for massive}) and the quantum 
corrections coming from the renormalization group. 
The expansion of the homogeneous, isotropic universe means 
a conformal transformation of the metric 
$\,g_{\mu\nu}(t) = {\bar g}_{\mu\nu} \exp\,[\si(\eta)]$.
On the other hand, the renormalization group in curved
space-time corresponds to the scale transformation of the 
metric $\,g_{\mu\nu} \to g_{\mu\nu}\cdot e^{-2\tau}\,$ 
simultaneous with the inverse transformation of all 
dimensional quantities \cite{book}. For any $\mu$ we 
have $\mu \to \mu\cdot e^{\tau}$. One can compare the 
$\,\sigma\,$-dependence of the anomaly-induced effective 
action (\ref{quantum for massive}) and the
$\,\tau$-dependence 
of the renormalization-group improved classical action
\beq
S_{vacuum}[P(\tau)]\,,
\label{RGEA}
\eeq
where $\,P=\left(a_{1,2,3,4},G,\La\right)\,$ denote 
vacuum parameters of the theory and $P(\tau)=P_0 + \beta_P\tau$.
The expression (\ref{RGEA}) is a leading-log
approximation for the solution of the renormalization 
group equation for the effective action \cite{book}
\begin{eqnarray}
\Gamma[e^{-2\tau}g_{\alpha\beta},{\Phi_i},P,\mu ] =
\Gamma[g_{\alpha\beta},{\Phi_i}(\tau),P(\tau),\mu ]\,,
\label{RG}
\end{eqnarray}
where ${\Phi_i}$ are matter fields. 
It is easy to see that (\ref{quantum for massive}) becomes
completely equivalent to (\ref{RGEA}) if we set $\sigma=const$.
The coefficient $f$ is a factor of the $\beta$-function
for the inverse Newton constant $\,1/16\pi G$ and the 
coefficient $g$ is a factor of the $\beta$-function for the 
cosmological term $\La/8\pi G$. Indeed, (\ref{RG}) can be 
considered as a generalization of the renormalization group 
improved classical action (\ref{RGEA}). After all, one can 
consider Eq. (\ref{quantum for massive}) as a leading-log 
approximation for the effective action for the massive fields.
This approximation 
picks up the logarithmic quantum corrections and is reliable 
in the high energy region where masses of the fields are much 
smaller than the Hubble constant $H$. The anomaly-induced 
expression may be even seen as a $\overline{\rm MS}$-scheme
based local version of renormalization group \cite{fossil}, 
but the consistent formulation of such local renormalization 
group is not available yet.

\section{Applications of the anomaly-induced action}

\qquad
The most important areas of application for the quantum 
corrections to the action of vacuum are cosmology and the 
black-hole physics. In both cases the anomaly-induced
effective action described in the previous section 
is behind the most well-established results, such as 
Hawking radiation and Starobinsky model of inflation. 
In this part of our review we shall give a very 
brief account of these two subjects. We shall 
follow the original publications \cite{balsan,anju,asta} 
where one can find more detailed considerations. 

\subsection{Vacuum states in the vicinity of a black hole}

\qquad
At the classical level, the black hole does not emit 
radiation, but such emission can take place if we 
take quantum effects into account. 
After the theoretical discovery of the black hole 
evaporation by Hawking \cite{Hawking75}, the same 
result has been obtained from analytical estimates 
of $\langle T_{\mu\nu}\rangle$ for matter fields 
propagating in a fixed static Schwarzschild black 
hole geometry. Indeed, the black hole evaporation 
phenomenon has not been confirmed experimentally. 
In the situation when the complete information on the 
vacuum state or, equivalently, on the effective action 
of vacuum is unavailable, one can not be absolutely 
certain that such evaporation really takes place. 
However, all known 
methods converge predicting such quantum radiation.  
The completely consistent approach would involve the 
simultaneous solution of the following two problems: 
solving equations of motion derived from the effective 
action and derivation of the quantum evaporation of 
the black hole. However, the practical calculations 
which have been performed so far, are based on the 
reduced approach, when the quantum effects are 
estimated on the background of the purely classical 
solution. Here we shall use the same approximation 
\cite{balsan}.

The idea to derive the Hawking effect starting from 
the effective action of vacuum is natural.
The main problem here is to choose the appropriate 
method for evaluating the effective action. For 
example, in \cite{MWZ} (see also \cite{balsan-2d})
the starting expression has been derived starting 
from the two-dimensional conformal anomaly in the 
dimensionally reduced (from $d=4$) theory. As a result 
there are serious difficulties and the fit to the standard 
result requested some artificial procedure related 
with introducing an effective potential of the 
conformal factor of the metric. Similar situation 
takes place in recent works \cite{Vilk07}, where
the consideration was based on a new complicated 
approach to deriving the effective action of vacuum. 

Let us notice that no one of existing techniques for 
the effective action of vacuum is perfect in a sense 
there is always an ambiguity. The unique case when 
we can arrive at the exact one-loop result is the 
conformally trivial (e.g., FRW) metric, but this is 
definitely different from the black-hole situation. 
Then, since it is known the Hawking effect is closely 
related to the four-dimensional conformal anomaly 
\cite{ChrFull}, it 
is natural to expect the best fit with the traditional 
methods if starting from the anomaly-induced effective 
action. Indeed, this approach really enables one to 
classify the vacuum states in a somehow clear way and 
reproduce the standard result for the Hawking 
radiation \cite{balsan}. Moreover, the analogy with 
the black hole vacuum turns out to be very fruitful 
for the investigation of the gravitational waves in a 
anomaly-induced background \cite{anju}, where we achieve 
a close fit with the well known result of \cite{star} 
and with the consequent calculations \cite{hhr}. 

Consider the application of the anomaly-induced 
effective action (\ref{finaction}) to the analysis of 
the quantum effects in the vicinity of a black hole. 
We shall see that the different vacuum states of the 
black hole (Boulware,
Hartle-Hawking and Unruh) correspond to the different 
choices of initial conditions for the two auxiliary fields
\cite{balsan}. The result can be generalized for the 
Reissner-Nordstrom spacetime case \cite{Reis-Nord}.
Let us stress that such classification can not be
accomplished by using only one field $\ph$. Therefore the
correspondence with other approaches to Hawking radiation
confirms that our considerations about introducing the 
second auxiliary scalar $\,\psi\,$ are correct.

Detailed analytical and numerical investigations, 
based on the analysis of $\langle T_{\mu\nu}\rangle$ 
in the classical black hole background have been 
performed in \cite{molti}. One of the fundamental
properties of this background is the existence of 
three different vacuum quantum states, which are 
defined as follows: 
\vskip 1mm

i) The Boulware vacuum $|B\rangle$ \cite{bula} is 
obtained by choosing {\it in} and {\it out} modes to 
be positive frequency with respect to the Killing 
vector $\partial_t$ of the Schwarzschild metric 
(\ref{b3}). 
This state reproduce the Minkowski vacuum $|M\rangle$ 
asymptotically, in the limit $r \to \infty$ we find
$\,\langle B|T_{\mu\nu}|B\rangle \sim 1/r^6$. 
At the same time the behavior of the same average 
value on the horizon $r=2M$ is divergent in a free 
falling frame. In this situation, for the classical 
observables, one can choose another coordinates, e.g. 
(\ref{coed}), such that the horizon becomes free of the 
singularities. In quantum theory this corresponds to 
choosing another vacuum state. 
\vskip 1mm

ii) For the Unruh vacuum $|U\rangle$ \cite{uuru} the 
{\it in} modes have positive frequency with respect 
to $\partial_t$, while the {\it out} modes are taken to 
be positive frequency with respect to Kruskal's 
$U=-4Me^{-u/4M}$ (\ref{coed}). The value 
$\langle U|T_{\mu\nu}|U\rangle$ is regular on the 
future event horizon but not on the past one. 
Asymptotically in the future 
$\langle U|T_{\mu\nu}|U\rangle$ has the form of a 
flux of radiation at the Hawking temperature 
$T_H=1/8\pi M$. Hence this vacuum state 
is the most appropriate to discuss evaporation of 
black holes formed by gravitational collapse of 
matter. 
\vskip 1mm

iii) The Israel-Hartle-Hawking $|H\rangle$ vacuum 
state \cite{ishh} is obtained by choosing {\it in} 
modes to be positive frequency with respect to the 
Kruskal's coordinate $V=4Me^{v/4M}$ and {\it out} 
modes positive frequency with respect to the affine 
parameter along the past horizon. 
$\,\langle H|T_{\mu\nu}|H\rangle\,$  for $r\to\infty$ 
describes a thermal bath of radiation at the Hawking 
temperature $\,T_H$. 
This vacuum state corresponds to a black hole in 
equilibrium with a surrounding thermal bath.
Asymptotically both in the future and the past 
$|H\rangle \neq |M\rangle$. 
\vskip 2mm

The existence of three distinct vacuum states in quantum 
theory looks natural, in sense they reflect distinct 
choices of coordinates systems and construction of 
different {\it in} and {\it out} modes with respect 
to the corresponding coordinates. The main difference 
between classical and quantum theories is that, in 
the first case we know how to transform the coordinates. 
So, the natural question is how to perform the transition 
between different vacuum states $|H\rangle$, $|B\rangle$
and $|U\rangle$? 

The anomaly-induced part of the effective action do not 
make any reference to a particular quantum state. Therefore 
one has to look for the non-universal part of effective 
action in order to implement one or another state. 
We remember that the solution (\ref{nonloc}) is exact 
only for the conformally-trivial metric, such as the 
Friedmann-Robertson-Walker (FRW) one. 
In all other cases the conformal invariant functional 
$S_c[g_{\mu\nu}]$ is a source of uncertainty. 
Furthermore, in order to apply the 
quantum correction (\ref{finaction}) one has to fix
the extended set of boundary conditions, including the 
ones for the auxiliary scalar fields $\ph$ and 
$\psi$. It turns out that the identification of the 
Boulware and Unruh vacuum states can be performed, in 
the leading approximation, by choosing an appropriate 
initial and boundary conditions for the auxiliary 
scalars. In order to achieve the Israel-Hartle-Hawking
one has to make an additional adjustment of the 
conformal functional. We will not go into details 
here, because one can easily find them in \cite{balsan}
(see also \cite{balsan-2d} where the more simple 
effective two-dimensional case has been treated) 
and present just the list of main results. 

The equations of motion for the auxiliary fields 
$\ph$ and $\psi$ have the form
\beq
-\,\frac{1}{\sqrt{-g}}\frac{\delta {\bar \Ga}}{\delta\phi}
&=&
\Delta_4\phi + k_1 C^2 + k_2(E-\frac{2}{3}\Box R)=0\,,
\nonumber
\\
\frac{1}{\sqrt{-g}}\frac{\de {\bar \Ga}}{\delta\psi}
&=& - \Delta_4\psi + l_1 C^2 = 0\,,
\label{15}
\eeq
where we used the notations (\ref{kl}). In the Ricci-flat 
space the operator $\De_4$ boils down to $\Box^2$, such that 
we get 
\beq
\Box^2\phi = \frac{\al\,(GM)^2}{r^6}\ ,
\label{emsb}
\eeq
where $\,\alpha = -48(k_1+k_2)$. The equation for $\psi$ 
can be obtained by trading $\al \to \be=48l_1$. Until the 
end of this subsection we shall set the Newton constant 
$\,G=1$. It proves useful
to introduce the traceless tensor $K_{\mu\nu}$ (an 
explicit form of this object can be found in \cite{balsan})
\beq
K_{\mu\nu}(\ph)\,=\,
-\,\frac{1}{\sqrt{-g}}\,\frac{\de}{\de g^{\mu\nu}}
\int d^4x\sqrt{-g}\, \ph\De_4\ph \,,
\label{17}
\eeq
then 
\beq
- \frac{2}{\sqrt{-g}}\frac{\de \Ga}{\de g^{\mu\nu}}
&=& \,\langle S_{\mu\nu} \rangle \,\,=\,\,
K_{\mu\nu}(\ph)-K_{\mu\nu}(\psi) 
-  8\nabla^{\lambda}\nabla^{\tau}Z R_{\mu\lambda\nu\tau} 
+ g_{\mu\nu}\, Z R^2_{\rho\sigma\alpha\beta} 
\nonumber
\\
&-& 
4Z\,R_{\mu\rho\lambda\tau}\,{R_{\nu }}^{\rho\lambda\tau} 
- \frac{4k_2}{3}\,
\left[(\nabla_\mu\nabla_\nu {\Box} \ph)
- g_{\mu\nu} ({\Box}^2 \ph) \right] + ... \,,
\label{18}
\eeq
where we omitted those term which vanish on the Ricci flat 
background such as the Schwarzschild metric. Let us notice 
that $\langle S_{\mu\nu} \rangle$ represents only that 
part of the average Energy-Momentum Tensor which can be 
restored from anomaly, being in general different from 
the unknown complete quantity $\langle T_{\mu\nu} \rangle$. 

The strategy for identifying the given vacuum state is as 
follows. One solves the equations (\ref{15}) for the 
auxiliary fields and replace the solutions into (\ref{18}). 
These solutions will always depend on the set of integration 
constants and one can try to find the ``correct'' ones to 
identify 
$\,\langle V|T_{\mu\nu}|V\rangle\,$ for the given 
vacuum state $\,|V\rangle$, being it 
$\,|V\rangle=|B\rangle$, \ $|U\rangle$ \ or \ $|H\rangle$. 

The general solution for $\ph$ has the form 
$\,\,\phi(r,t)=d\cdot t + w(r)\,$,
where $\,w(r)\,$ satisfies the equation
\beq
\frac{dw}{dr} 
&=& \frac{B}{3}r +\frac{2MB}{3}-\frac{A}{6}-\frac{\alpha}{72M}
+ \frac{1}{r-2M}
\left(\frac{4}{3}BM^2 + \frac{C}{2M}-AM -\frac{\alpha}{24}\right)
\nonumber \\
&-&
\frac{C}{2M}\frac{1}{r} 
+ \ln r \left[ -\frac{\alpha M}{18}\frac{1}{r(r-2M)}
- \frac{r^2}{3(r-2M)}
\left(\frac{A}{2M}-\frac{\alpha}{48M^2}\right)\right]
\nonumber \\
&+&  \frac{\left(r^3 -8M^3\right)\ln(r-2M)}{3r(r-2M)}\, 
\left(\frac{A}{2M}-\frac{\alpha}{48M^2}\right)
\label{22}
\eeq
and $(d,A,B,C)$ are constants that specify the homogeneous 
solution $\Box^2\phi=0$ and hence the quantum state. For 
$\psi$ we have a similar solution, but with another set of 
integration constants $(d',A',B',C')$. It is important 
that these two sets are independent on each other, due to 
the independence of $\,\ph\,$ and $\,\psi$. 

In the case of a Boulware state $|B\rangle$ we request 
$|B\rangle \to |M\rangle$ when $r \to \infty$. 
For the auxiliary scalars in the Minkowski vacuum we
can safely set $\ph=\psi=0$. This asymptotic conditions
and also cancelation of singular terms in the solution 
of (\ref{22}) enables one to fix all constants and we 
arrive at the asymptotic expressions \cite{balsan}
\beq
\label{ciuco}
\langle B| S_{\mu}^{\nu}  |B \rangle 
\to  
\frac{1}{2} 
\frac{\alpha^2-\beta^2}{(24)^2}\,
\frac{1}{(2M)^4\,\left(1-2M/r\right)^2}\,
\left( 
\begin{array}{cccc}
-1 & 0 & 0 & 0  \\ 
0 & 1/3 & 0 & 0 \\ 
0 & 0 & 1/3 & 0 \\ 
0 & 0 & 0 & 1/3 \end{array}
\right)
\eeq
for \ $r\to 2M$ \ and
\beq
\langle B| S_{\mu}^{\nu}  |B \rangle 
\propto {\cal O}\big(r^{-6}\big)
\qquad \mbox{for} \qquad  r \to \infty\,.
\label{calo}
\eeq
This behaviors perfectly reproduce the one which is 
observed in the framework of other methods \cite{molti}. 
\ Namely, we have ${\cal O}\big(r^{-6}\big)$
in the space infinity and the quadratic divergence 
${\cal O}\big(\left(1-2M/r\right)^{-2}\big)$ near the 
horizon in the Boulware case.
\vskip 1mm

Another success of our method comes in the case of 
the Unruh vacuum. Using the same treatment of auxiliary 
fields, but choosing another values of the constants 
$(d,A,B,C)$ and $(d',A',B',C')$, we meet, near the 
horizon, the following asymptotic behavior \cite{balsan}:
\beq
\langle U| S_{a}^{\ b}|U \rangle 
\sim  
\frac{\alpha^2-\beta^2}{2(48M^2)^2} 
\left( 
\begin{array}{cc}
1/f & -1 \\ 1/f^2 & -1/f 
\end{array}\right)\,,
\qquad \ r\to 2M\,,\quad\mbox{where}\quad a,b=r,t\,,
\label{28}
\eeq
which is regular on the future horizon. The asymptotic 
form at the space infinity is 
\beq
\langle U|  S_{\mu}^{\ \nu} |U  \rangle \to  
\frac{\alpha^2-\beta^2}{2r^2(24M)^2} 
\,\left( 
\begin{array}{cccc}
-1 & -1 & 0 & 0 \\ 
1 & 1 & 0 & 0   \\ 
0 & 0 & 0 & 0   \\ 
0 & 0 & 0 & 0
\end{array}\right)\,,
\quad \ r\to \infty \ .
\label{giango}
\eeq
The results of eqs. (\ref{28}), (\ref{giango}) are 
in exact agreement with the standard ones on 
Hawking radiation \cite{molti,dewitt-75}, once the 
luminosity $L$ of the radiating black hole is identified 
with 
\beq
\frac{L}{4\pi} =\frac{(\alpha^2-\beta^2)}{2(24M)^2}\ .
\label{lume}
\eeq

A little bit more complicated situation takes place for 
the Hartle-Hawking vacuum, where one should not only 
properly choose the initial conditions but also fine-tune 
the coefficient $l_1$ to order to achieve correspondence 
with the results achieved by other methods. From the 
general perspective this situation looks somehow natural 
because, as we already explained earlier, the unknown 
conformal invariant functional $S_c$ is relevant for 
the spherically symmetric metric. The modification of 
$l_1$ is nothing else but the adjustment of $S_c$. So, 
we can consider as a kind of luck that the most 
essential part of the classification of the vacuum 
states can be performed by the use of the initial 
conditions for $\,\ph\,$ and $\,\psi$, such that only 
in the case of $|H\rangle$ we are forced to modify the 
conformal term. 

\subsection{Modified Starobinsky model.} 

\qquad
Another interesting application of the 
anomaly-induced action is the Modified Starobinsky Model
or anomaly-induced inflation \cite{anju,Shocom,asta}.
Let us follow \cite{Shocom,asta} and first consider 
quantum effects of massless fields, that is use the 
effective action (\ref{quantum}). At the second stage 
we shall take masses of the fields into account, that 
means we shall use the action (\ref{quantum for massive}). 

As a first step consider an empty space, when matter is 
absent. There are two equivalent ways to arrive at the 
cosmological solution of the theory with quantum 
corrections: \ using the $\,(0$-$0)$-component 
\cite{fhh,star} or via the anomaly-induced effective 
action \cite{buodsh,book}. Let us choose the last option. 
The resulting equation has, for $\,k=0$ FRW metric, 
the following form\footnote{The cases $k=\pm 1$ are 
quite similar \cite{asta} and we do not consider them 
here.}:
\beq  
\frac{{\stackrel{....}{a}}}{a}
+\frac{{3\stackrel{.}{a}} {\stackrel{...}{a}}}{a^2}
+\frac{{\stackrel{..}{a}}^{2}}{a^{2}}
-\left( 5+\frac{4b}{c}\right) 
\frac{{\stackrel{..}{a}} {\stackrel{.}{a}}^{2}}{a^3}
-\frac{M_{P}^{2}}{8\pi c}
\left( \frac{{\stackrel{..}{a}}}{a}+
\frac{{\stackrel{.}{a}}^{2}}{a^{2}}
-\frac{2\Lambda }{3}\right)\,=\,0\,,
\label{foe}
\eeq
where the coefficients $\,b,\,\,c\,$ are defined in 
 (\ref{abc}). The last equation does 
not depend on the coefficient $\,\om$, because the 
Weyl tensor vanish for the FRW background. 
The equation (\ref{foe}) has a remarkable particular 
solution
\beq
a(t) \,=\, a_0 \cdot \exp(Ht)
\label{flat solution}
\eeq
where 
\beq
H\,=\, \frac{M_P}{\sqrt{-32\pi b}}\,\left(1\pm 
\sqrt{1+\frac{64\pi b}{3}\frac{\Lambda }{M_P^2}}\right)^{1/2}.
\label{H}
\eeq
As far as $\,\La \ll M_P^2$, we meet two very different 
values of $H$ (we consider $\La > 0$)
\beq
H_{c}\,\approx\,\sqrt{\frac{\Lambda }{3}}
\,\,\,\,\,\,\,\,\,\,\,\,
{\rm and}\,\,\,\,\,\,\,\,\,\,\,\,
H_S\,\approx\,\frac{M_P}{\sqrt{-16\pi b}}\,.
\label{HH}
\eeq
The solution with $H_c$ is the one of the theory without 
quantum corrections. The second value $\,H_S\,$ corresponds 
to the inflationary solution of Starobinsky \cite{star}. 
Let us notice that the sign of $\,b\,$ is negative 
independent on the particle content (\ref{abc}). 
Let us remark that the 
expression ``particle content'' here corresponds to the 
degrees of freedom contributing to the vacuum effective 
action in the virtual loops and has nothing to do with the 
matter content of the universe. 

Further understanding of the inflationary solution 
requires analyzing its stability properties.  
The phase portrait of the theory may look very different 
depending on the sign of the coefficient $\,c$, that is on 
the coefficient of the local $\,\int\sqrt{-g}R^2$-term
\cite{star,asta}. 
The inflationary solution is stable for a positive $\,c\,$ 
and is unstable in the case $\,c<0$. In the last case there 
are several stable points (attractors),
one of which corresponds to the usual 
\ $a \sim t^{1/2}$ \ solution. It is interesting that the 
existence of the exponential solution is possible due to 
the non-local and universal ${\Ga}_b$-term in the 
anomaly-induced effective action, while the stability 
depends on the local  $\,\int\sqrt{-g}R^2$-term only.

The Starobinsky model looks appealing, in particular 
because it has a solid QFT background and there is no 
need to introduce a special inflaton field. The original 
model \cite{star} is based on the unstable solution $\,c<0$.
In this case one has to choose the initial conditions in 
a very special way. First of all, the initial point must 
be very close to the exact exponential solution 
(\ref{flat solution}), 
such that the inflation lasts long enough. Moreover, one
has to provide that, after
the inflationary phase ends, the Universe will approach 
the attractor corresponding to the FRW solution, and not 
to other, physically unacceptable, attractors. All the 
matter content of the Universe is created after the 
inflation ends through the decay of the massive degree of 
freedom induced by anomaly. Many interesting aspects of 
this model, including gravitational waves and problem of 
singularities, has been investigated starting from the 
papers \cite{star,vile,Ander}. The first investigation 
of inflationary density perturbations has been also 
performed for this model \cite{much}. 

An alternative way to use the solution (\ref{quantum}) 
\cite{anju} is based on a positive value of $\,c\,$ and 
therefore on the stability of the exponential solution 
at the beginning of inflation. It turns out that the 
stable inflation is very robust with respect to the choice 
of initial data \cite{asta}, providing the following 
two advantages: \
First, one does not need to consider an empty universe
in the initial state, since the stable inflation ``kills'' 
any matter content in a few Planck times \cite{asta-OPC}. 
\ Second, there is no problem with the 
unphysical solutions of the ``run-away'' sort, the
measure of the corresponding initial data is infinitesimal 
for $\,c>0$. 

The real problem of the stable $\,c>0\,$ inflationary 
model is to understand the period when the inflation 
ends. If we stay within the original framework 
\cite{fhh,star}, that is consider only massless conformal 
fields, the stable inflation will be eternal. The modified
Starobinsky model solves this problem using the effective 
quantum field theory approach \cite{Shocom} and in 
particular the notion of decoupling. 
Using the relations (\ref{abc}), the condition of stability
$\,c>0\,$ can be cast into the form
\beq
N_v \,<\,\frac13\,N_{f}\,+\,\frac{1}{18}\,N_{s}\,.
\label{condition}
\eeq
The inflation is stable if there are many 
scalars and fermions, for a given amount of vectors. 
We know that in QFT vectors are responsible for the 
fundamental interactions. It is easy to see that the 
inequality (\ref{condition}) is not satisfied for the 
MSM with $\,N_{v,\,f,\,s}=(12,24,4)$. However, it 
is satisfied for the MSSM with $\,N_{v,\,f,\,s}=(12,32,104)$.
The same must be true for any realistic supersymmetric
model, because the supersymmetrization of the realistic 
model implies transformation of fermions and scalars 
into chiral superfields and vectors into vector 
superfields. Both operations require adding many spinor
and scalar superpartners (sparticles) while the vector 
sector of the theory remains the same. 

The transition between stable and unstable inflation 
can be associated with the soft SUSY breaking. Within 
this approach the sparticles are supposed to be very 
heavy compared to the usual particles. The inflation 
becomes unstable when the typical energy becomes smaller 
than the mass of the sparticles and they decouple. 
As before, we associate the typical energy with the 
Hubble parameter $\,H$. Let us denote $\,M_*\,$ the 
energy scale where the 
inequality (\ref{condition}) changes its sign to the 
opposite. The value $\,H_S\,$ is now seen just as 
initial value of $\,H\,$ and $\,H=M_*\,$ is the final 
point of the stable inflation. The transition occurs at 
the instant $\,t_f\,$ which 
is defined as a solution of the equation $\,H(t_f)=M_*$. 

The details of decoupling will be considered in the next 
section. The next problem is to find why the inflation 
slows down. In fact, this can be discovered at once if 
we take the masses of the fields
into account and use the result (\ref{quantum M}). The 
equation of motion for $a(t)$ has the form 
\beq  
\frac{{\stackrel{....}{a}}}{a}
\,+\, 
3\,\frac{\,{\stackrel{.}{a}}}{a}\,\frac{{\stackrel{...}{a}}}{a}
+ \frac{\,{\stackrel{..}{a}}^{2}}{a^{2}}
\,-\Big( 5+\frac{4b}{c}\Big)\, \frac{\,{\stackrel{..}{a}}}{a}
\,\frac{{\stackrel{.}{a}}^{2}}{a^{2}}
&-& 
\frac{M_P^2}{8\pi c}
\,\Big[\,\Big(\,\frac{{\stackrel{..}{a}}}{a}
+ \frac{\dot{a}^2}{a^2}\Big)
\cdot (1-\tilde{f}\cdot{\rm ln} a)
\,-\,\frac{\tilde{f}}{2}\,\frac{\dot{a}^2}{a^2}\,\Big]
\nonumber
\\
&+& 
\frac{M_P^2\La}{12\pi c}
\,(\,1 - \tilde{g}\cdot \ln a - \tilde{g}/4\,)\,=\,0\,.
\label{central a}
\eeq
where we introduced useful notations
\beq
\tilde{f} = \frac{16\pi f}{M_P^2} = 
\frac{1}{3\pi}\,\sum_{f}\,\frac{N_f\,m_f^2}{M_P^2} 
\,;\,\,\,\,\,\,\,\,\,\,\,\,\,\,\, 
\tilde{g} = \frac{8\pi g}{M_P^2\La}
\,=\,\frac{1}{4\pi}\,\sum_{s}\,\frac{N_s\,m_s^4}{M_P^2\La}
\,-\,\frac{1}{\pi}\,\sum_{f}\,\frac{N_f\,m_f^4}{M_P^2\La}\,.
\label{replace11}
\eeq
The leading terms in the eq.
(\ref{central a}) are the same as in the massless case
(\ref{foe}), but the Planck mass and the cosmological 
constant to be replaced by the variable expressions
\beq
M^2_P\,\to\,M^2_P\,(1-\tilde{f}\ln a)\,,
\label{central sigma}
\\
\La\,M^2_P\,\to\,\La\,M^2_P\,(1 - \tilde{g}\ln a)\,.
\label{central sigma C}
\eeq
Let us assume that, because of 
supersymmetry, the quantum loops contributions to the  
cosmological constant do cancel \ $\tilde{g}\approx 0$
\ \cite{asta}.
Furthermore, \ $\La$ \ itself is negligible at the energy  
scale of inflation. Then the solution 
of equation (\ref{central a}) can be obtained by 
solving (\ref{central sigma}) in the form
\beq
\si(t)\,=\,\ln a(t)
\,=\,H_0\,t\,-\,\frac{H^2_0}{4}\,\tilde{f}\,t^2\,.
\label{parabola}
\eeq
The numerical analysis confirms the parabolic dependence 
(\ref{parabola}) with excellent precision \cite{asta}. 
The total number of the inflationary 
$\,e$-folds for different models of the SUSY breaking
can be obtained from the relation (\ref{parabola}).
For example, in the case of MSSM we have $M_*\sim 1\,TeV$, 
then $\,\tilde{f}\sim (M_*/M_P)^2 = 10^{-32}$ and 
the total amount of inflationary $\,e$-folds is $10^{32}$.

One can use the effective QFT approach to perform additional
tests of quantum corrections (\ref{quantum}). If we believe 
to the decoupling idea, the present-day Universe looks like 
a perfect object to apply the effective approach. In the present-day Universe only the photon should be treated as 
an active physical degree of freedom and hence, according 
to (\ref{condition}), $\,c<0$. Then the explicit analysis 
of stability performed in \cite{asta}, shows that the 
solution $\,H_{c}\,$ in eq. (\ref{HH}) is stable if $\La>0$.
Furthermore, an additional advantage of the anomaly-induced 
is the moderate behavior of the gravitational waves or, in 
other words, restricted production of gravitons \cite{anju}. 

Until 
now we could see only good points of the anomaly-induced 
inflation. This model is completely based on the QFT results, 
it does not require introducing new artificial entities such 
as inflaton, does not require fine tuning for the action or 
for initial data, moreover the gravitational waves do not 
grow too much during the stable phase, making consistent 
the whole approach based on the homogeneous and isotropic 
metric. Unfortunately, there is also a real problem with 
this model. As we already know, the total 
amount of the inflationary $e$-folds is enormous. 
However, only the last $60-70$ of these $e$-folds 
are physically relevant. In order to extract the physical 
information one needs, therefore, an effective action
in the region 
when $H\propto M_*$. However, as we have already discussed 
above, we have no approximation for this situation, 
which lies between the light-mass and the heavy-mass 
extremes. No model exists for the intermediate situation. 
Until such model will be constructed, the only way to 
study inflation is through the inflaton approach. 
However, the very fact that the inflation can be explained 
by quantum effects of matter fields surely looks 
interesting and greatly increase the importance of the 
further study of effective action of massive fields.

\section{Effective action for massive fields}

\qquad
For a while we have seen that two kinds of information 
are available about the effective action in curved space. 
First, we can completely calculate the divergences and 
remove them using the renormalization procedure. The 
procedure works perfect in curved space-time, independent 
on that we do not have access to the complete expression 
for the effective action or, equivalently, can not obtain 
the vacuum state of the quantum fields on a classical 
gravitational 
background. The key point is that the divergences are 
given by the well known local expressions and the unknown 
part of the 
effective action is non-local. Second, we are able to 
obtain the non-local part within the anomaly-induced scheme. 
In this section we shall go further and obtain, in a regular 
way, some non-local terms for the massive fields theory. 
We shall present only the main results here, the details 
can be found in the original papers \cite{apco,bexi}. 


\begin{quotation}
\begin{figure}
\qquad
\qquad
\quad
\includegraphics[width=0.65\textwidth]{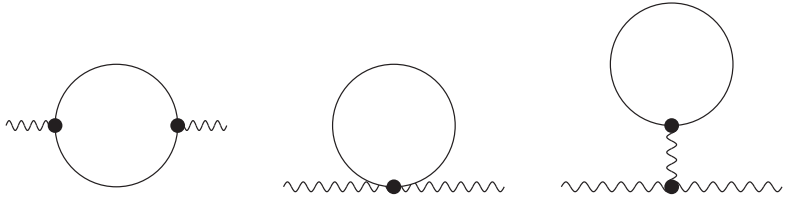}
\\
\caption{\sl
The  one-loop diagrams contributing to the 
polarization operator of gravitons. The matter 
field loop is connected to external gravitational 
fields represented by wavy lines.}
\end{figure}
\end{quotation}

\subsection{Decoupling theorem for gravity}

\qquad
As we learned above, the vacuum divergences can be 
removed by the renormalization of the action 
(\ref{vacuum}). This procedure leads to the 
$\,\overline{\rm MS}$ scheme - based
renormalization group running, which enables one to 
restore the high energy asymptotics of the effective 
action by trading $\,\ln \mu^2\,$ by $\,\ln \Box\,$  
\cite{buwolf,book}. Now we ask the following questions: 
are these $\,\ln \Box\,$ terms really there? And what 
happens at low energies, when the $\,\overline{\rm MS}\,$
scheme does not work? We know that in QED the main 
phenomenon at the low-energy regime (we shall call 
it IR) is the decoupling of massive fields \cite{AC}.
One can expect something like that in the vacuum 
gravitational sector, because the general structure 
of the Feynman diagrams is similar (see, e.g., Fig. 2). 

The most obvious way to the decoupling theorem is to 
derive the effective actions and the $\,\beta$-functions
$\,\be_{\la},\,\be_R,\,\be_{1,..,4}\,$ through the
\ $h_{\mu\nu}=g_{\mu\nu}-\eta_{\mu\nu}\,$ expansion 
approach \cite{apco}. This method opens the way for the 
momentum-subtraction renormalization scheme, that is  
the standard approach for the Appelquist and Carazzone 
theorem \cite{AC}. 
A useful alternative to the flat-space expansion is
a resummation of the Schwinger-DeWitt series,
culminated in the heat kernel solution of \cite{bavi90}. 
The results obtained within these two methods coincide
\cite{apco}, 
and for the scalar field we arrive at the following 
result in the ${\cal O}(R^2_{...})$ approximation:
\beq
{\bar \Ga}^{(1)}_{scalar}
&=& \frac{1}{2(4\pi)^2}\,\int d^4x \,\sqrt{-g}\,
\Big\{\,\frac{m^4}{2}\cdot\Big(\frac{1}{\ep}
+\frac32\Big)\,+
\,\Big(\xi-\frac16\Big)\,m^2R\,
\Big(\frac{1}{\ep}+1\Big)
\nonumber
\\
&+& \frac12\,C_{\mu\nu\al\be} 
\,\Big[\frac{1}{60\,\ep}+k_W(a)\Big] C^{\mu\nu\al\be}
\,+\,R \,\Big[\,\frac{1}{2\ep}\,\Big(\xi-\frac16\Big)^2\,
+ k_R(a)\,\Big]\,R\,\Big\}\,.
\label{final}
\eeq
Here $\,\ep\,$ is the parameter of dimensional regularization
$$
\frac{1}{\ep}=\frac{2}{4-n}
+\ln \Big(\frac{4\pi \mu^2}{m^2}\Big) - \ga\,.
$$
It is easy to see that the divergences are exactly the 
same as in the $\,\overline{\rm MS}$ scheme (\ref{tot}),
(\ref{abc}). The qualitatively new part is encoded into 
the formfactors (nonlocal insertions produced by quantum 
corrections), which have the form
\beq
k_W(a) &=& \frac{8A}{15\,a^4}
\,+\,\frac{2}{45\,a^2}\,+\,\frac{1}{150}\,,
\nonumber
\\
k_R(a) &=& 
A\Big(\xi-\frac16\Big)^2-\frac{A}{6}\,\Big(\xi-\frac16\Big)
+\frac{2A}{3a^2}\,\Big(\xi-\frac16\Big)
+\frac{A}{9a^4}-\frac{A}{18a^2}+\frac{A}{144}+
\nonumber
\\
&+& \frac{1}{108\,a^2}
-\frac{7}{2160} + \frac{1}{18}\,\Big(\xi-\frac16\Big)\,.
\label{W}
\eeq
Here we used notations 
\beq
A\,=\,1-\frac{1}{a}\ln\,\Big(\frac{2+a}{2-a}\Big)\,,
\qquad a^2 = \frac{4\Box}{\Box - 4m^2}\,. 
\label{Aa}
\eeq
Similar expressions can be found for massive fermion and
vector loops \cite{apco} and also for the scalar field 
background, where the formfactors can be obtained for the 
$\,\ph^2 R\,$ and $\,\ph^4\,$ terms. 
In principle, there is a possibility to obtain similar 
expressions in the third order in curvatures using the 
corresponding results for the heat kernel \cite{bavi90}. 

In the flat space the renormalization group scaling 
is defined as the momentum scaling 
\ $p^2 \to e^{2\tau} p^2.$ \
In curved space renormalization group may be defined 
as scaling of metric in the coordinate space
\ $g_{\mu\nu} \to e^{-2\tau} g_{\mu\nu}$, that is
consistent with the flat space limit because
\ $p^2=\eta_{\mu\nu}p^{\mu}p^{\nu}$.
In the mass-dependent scheme we define the $\be$-function
for the parameter $\la$, as
$$
\beta_\la\,=\, -\,2p^2\,\frac{\pa \la}{\pa p^2}\,,
$$
and furthermore identify $\,p^2\,$ with $\, -\Box$. 
For the Weyl 
term we have, following this procedure \cite{apco},
$$
\be_1\,=\, -\,\frac{1}{(4\pi)^2}\, \left(\, 
\frac{1}{18a^2}\,-\,\frac{1}{180}\,-\, 
\frac{a^2-4}{6a^4}\,A\,\right)\,.
$$
From this expression we obtain
\beq
\be_1^{UV} &=& - \frac{1}{(4\pi)^2}\frac{1}{120} 
+ {\cal O}\left(\frac{m^2}{p^2}\right)
=\be_1^{\overline {MS}}
+ {\cal O}\left(\frac{m^2}{p^2}\right)
\nonumber
\\
\mbox{and} \qquad
\be_1^{IR} &=& -\,\frac{1}{1680\,(4\pi)^2}\,\cdot\,
\frac{p^2}{m^2}
\,+\,{\cal O}\left(\frac{p^4}{m^4}\right)\,,
\nonumber
\eeq
that is a quadratic decoupling 
in agreement with the Appelquist and Carazzone theorem.
For $a_3$ the situation is similar. The UV regime gives 
the same result as in (\ref{abc}),
$$
\be_3^{UV}=-\frac{1}{180(4\pi)^2}
+{\cal O}\left(\frac{m^2}{p^2}\right)\,,
$$
while the IR limit shows the standard quadratic decoupling
$$
\be_3^{scalar, \,IR}\,\,=\,\,-\,\,
\frac{1}{1260\,(4\pi)^2}\,\,\frac{p^2}{m^2}
\,\,+\,\,{\cal O}\left(\frac{p^4}{m^4}\right)\,.
$$
An important feature of $\,\be_3\,$ is that it 
changes sign between UV and IR regimes in the models 
with broken SUSY  due to the decoupling of sparticles
\cite{apco}.
As we have seen in the previous section, this property 
is required for the modified Starobinsky model of 
inflation.

One can, also, establish the quadratic decoupling in 
directly in the formfactors. The UV limit corresponds 
to \ $a\to 2$ \ and the IR limit to \ $a\to 0$. It is 
easy to check that in the IR limit
 \ $A \sim -a^2/12 \to 0$ \ 
and therefore in the IR both $k_W(a)$ and $k_R(a)$ tend 
to zero as \ ${\cal O}\big(a^2\big)$. 
One can observe the same behavior for the formfactors 
in the scalar sector $\,k_\la(a)\,$ and $\,k_\xi(a)\,$ 
\cite{bexi}.  
The gravitational version of the Appelquist and Carazzone
theorem holds also in the case of spontaneous symmetry 
breaking \cite{sponta}, where the analysis is more 
cumbersome.

\subsection{Massless limit in the massive theory}

\qquad
As we have seen in the previous section, the dominating 
phenomenon in the massless theory is the conformal anomaly. 
It would be quite interesting to obtain anomaly or at least 
part of it from the second order in curvature expressions
such as (\ref{final}). For this end one has to consider the 
massless limit. 
Another interesting aspect of this limit in the 
specific case of the vector theory with the softly broken 
gauge invariance is the discontinuity of quantum 
corrections, which was recently described in details in 
\cite{BuGui}. Let us consider these two issues here. 

\subsubsection{Anomaly from the heat kernel solution.} 

\qquad
In the massless conformal limit $\,m^2=0,\,\,\xi=1/6\,$
we obtain, from the \ $R \left(k_R\right) R$-term, the 
expression 
\beq
-\,\frac{1}{12\cdot 180(4\pi)^2}\,
\int d^4x \,\sqrt{-g}\,R^2 \,,
\label{anom R}
\eeq
fitting perfectly with the conformal anomaly \cite{bavi3}. 
This result can be generalized, also, for massive fermions 
and vectors. Let us remark that the 
\ $R \left(k_R\right) R$-term remains non-local even for 
a very small but non-zero mass and, therefore, 
it becomes ambiguous only in the precisely massless 
theory. Let us remark that in the general nonconformal 
theory an infinite renormalization of the 
$\,\int\sqrt{-g}R^2$-term is always necessary to make the
theory finite. In the conformal 
massless case this term has to be included in order to 
fix the finite ambiguity. In this sense there is no 
\ $m\to 0$ \ discontinuity. 

It is also instructive to look at the massless limit
of the formfactor for the Weyl term
\beq
&\,& \left.\int d^4x\sqrt{-g}\,\,
C_{\mu\nu\al\be}\Big(
\frac{1}{60\,\ep}+k_W \Big)\,C^{\mu\nu\al\be}
\right|_{m\to 0}
\,=\,\qquad
\nonumber
\\
& = &
\frac{1}{60}\,\int d^4x\sqrt{-g}\,\,
C_{\mu\nu\al\be}\,\left\{\left[
\frac{2}{4-n}-\ln \Big(-\frac{\Box}{4\pi \mu^2}\Big)
\right] \,\,+\,\, const\right\}\,C^{\mu\nu\al\be}\,.
\label{massless}
\eeq
This term defines the leading-log quantum 
contribution to the propagation of the gravitational wave, 
that is the transverse traceless part of the gravitational 
perturbation \ $h_{\mu\nu}=g_{\mu\nu}- \eta_{\mu\nu}$.

It is easy to see that this expression, also, directly 
reproduce the Weyl-square term in the conformal anomaly. 
For this end we have to just use the definition 
(\ref{trace}) and the relation (\ref{deriv}). For the 
conformally transformed metric 
\ $g_{\mu\nu}={\bar g}_{\mu\nu}\,\exp\left(2\si\right)$,
we have  
$$
\sqrt{-g}\,C_{\mu\nu\al\be}\,C^{\mu\nu\al\be}
\,=\,\sqrt{-{\bar g}}\,{\bar C}_{\mu\nu\al\be}
\,{\bar C}^{\mu\nu\al\be}\,.
$$
The constant term in the {\it r.h.s.} of (\ref{massless})
is obviously irrelevant for the anomaly and the dependence 
on the conformal factor $\si$ may come 
only from 
$$
\ln \Big(-\frac{\Box}{4\pi \mu^2}\Big)
\,\,=\,\,2\si 
\,+\, \ln \Big(-\frac{{\bar \na}^2}{4\pi \mu^2}\Big)
\,\,+\,\,{\cal O}(\si^2)\,\,.
$$
Obviously, this gives the standard expression for the 
corresponding term in the anomaly after we apply the 
prescription (\ref{deriv}). 

The same approach can be 
applied to the massless but conformal non-invariant 
limit $\,m^2=0,\,\,\xi \neq 1/6\,$ in the 
\ $\int R \left(k_R\right) R$ term\footnote{A possible
cosmological implications of this action have been 
considered in \cite{espriu}.} 
\beq
&\,& \left.\int d^4x\sqrt{-g}\,\,\,
R\,\,\Big\{\frac{1}{2\ep}\,\Big(\xi-\frac16\Big)^2
\,+\, k_R(a)\Big\}\,R\,\,
\right|_{m\to 0}\,=
\nonumber
\\
& = &   \int d^4x\sqrt{-g}\,\,\,R\,\left\{\,
\frac{1}{2}\,\left[
\frac{2}{4-n}-\ln \Big(-\frac{\Box}{4\pi \mu^2}\Big)
\right]\,\Big(\xi-\frac16\Big)^2 
\,\,+\,\, \frac{1}{12\cdot 180(4\pi)^2} \right\}\,R\,.
\label{massless R}
\eeq
The last term here is nothing but (\ref{anom R}). 
It is remarkable that any deviation of $\xi$ from the 
precisely conformal value $1/6$ leads to the nonlocalities 
in the \ $\int R \left(k_R\right) R$-sector. 
If we keep in mind that the same also happens for any 
nonzero value of the mass, it becomes clear that 
the ambiguity of the $\Box R$ term in the anomaly or, 
equivalently, of the $\int R^2$-term in the effective 
action is something quite subtle. 

\subsubsection{Discontinuity of quantum corrections in the 
Proca theory}

\quad \
The next interesting issue is the massless limit for 
the quantum corrections in the theory with softly 
broken gauge symmetry. Two most important examples 
of such theories are photon and graviton with tiny 
masses. We shall follow the recent paper \cite{BuGui} 
but consider only the massive vector theory.

Our purpose is to compare 
quantum contributions of massless and massive photons 
to the vacuum effective action in curved space-time. 
Consider the Proca model described by the action 
\beq
S_P = \int d^4x \sqrt{-g}\,\Big\{
- \frac14\,F_{\mu\nu}^2 + \frac12\,M^2\,A_\mu^2\Big\}\,.
\label{Proca}
\eeq
The standard approach \cite{bavi85} gives the following 
expression for the one-loop contribution:
\beq
{\bar \Ga}^{(1)} \,=\,\frac{i}{2}\, 
\Tr\ln 
\left(\de_\al^\nu \Box - R_\al^\mu - M^2\de_\al^\mu\right) 
\,-\,\frac{i}{2}\,\Tr\ln \left(\Box - M^2\right) \,.
\label{Proca 2}
\eeq
In the massless limit the last term is twice the 
Faddeev-Popov ghost contribution, that is why we meet a 
discontinuity in this limit. An extra term is nothing 
but the contribution from some scalar field. 
In order to better understand the origin of this scalar, 
one can apply the 
St$\ddot{\rm u}$ckelberg procedure \cite{stuck} and 
consider a new action
\beq
S_P^\prime = \int d^4x \sqrt{-g}\,\Big\{
- \frac14\,F_{\mu\nu}^2 - \frac12\,M^2\,
\Big(A_\mu - \frac{1}{M}\,\pa_\mu \ph\Big)^2\Big\}\,.
\label{Proca mod}
\eeq
This expression possesses invariance under the 
gauge transformation
$$
A_\mu \to A^\prime_\mu = A_\mu + \pa_\mu \xi 
\qquad
\mbox{and}
\qquad
\ph \to \ph^\prime = \ph + \xi M\,.
$$ 
In the special 
gauge \ $\ph=0$ \ we come back to the action 
(\ref{Proca}). Concerning the quantum corrections, 
the gauge fixing dependence is irrelevant for the 
free fields actions such as (\ref{Proca}) and 
(\ref{Proca mod}). Hence, it is legitimate to use 
the most useful gauge fixing to evaluate the 
quantum contributions. 

Using the linear gauge fixing condition
\ $\chi = \na_\mu A^\mu - M\ph$, \ the sum of the 
action (\ref{Proca mod}) and the gauge fixing term,  \ 
$S_{gf}= - \frac12\,\int d^4x\sqrt{-g}\,\chi^2$, \ 
is cast into the form 
\beq
S^\prime = \frac12\,\int d^4x \sqrt{-g}\,\big\{ A^\al
\left(\de_\al^\nu \Box - R_\al^\mu 
- M^2\de_\al^\mu\right)A_\nu 
+ \ph\left(\Box - M^2\right)\ph \big\}\,.
\nonumber
\eeq
The  Faddeev-Popov ghost's operator has the form 
\ ${\hat H}_{gh} = \Box - M^2$, therefore the
effective action is 
\beq
\bar{\Ga}^{(1)} &=& 
\frac{i}{2}\,\Tr\ln \left(\de_\al^\nu \Box 
- R_\al^\mu - M^2\de_\al^\mu\right) 
\nonumber  
\\
&+&  \frac{i}{2}\,\Tr\ln \left(\Box - M^2\right) 
\,-\, i\Tr\ln \left(\Box - M^2\right) \,,
\label{Proca 5}
\eeq
that is nothing but (\ref{Proca 2}). An extra scalar was 
``hidden'' in the massive term of the vector. 

In order to understand the physical sense of the 
discontinuity of the massless limit, let us recall the 
result for the Proca field obtained in \cite{apco},
\beq
{\bar \Ga}^{(1)}_{vector}
&=&\frac{1}{2(4\pi)^2}\,\int d^4x \,\sqrt{-g}\,
\Big\{\,\frac{3}{2}\,M^4\cdot\Big(\frac{1}{\ep}
+\frac32\Big)\,+\,\frac{M^2}{2}\,R\,\Big(\frac{1}{\ep}+1\,\Big)
\nonumber
\\
&+& \frac12\,C_{\mu\nu\al\be} \,\Big[\,\frac{13}{60\,\ep} 
+ k^v_W(a)\,\Big] C^{\mu\nu\al\be}
\,+\,R \,\Big[\,\frac{1}{72\,\ep}\,+\, k^v_R(a)\,\Big]
\,R\,\Big\}\,.
\label{final-v}
\eeq
This expression is quite similar to the one for the 
scalar field (\ref{final}). 

The most interesting for us is the nonlocal finite  
formfactor
\beq
k^v_W(a) \,=\,
\,-\frac{91}{450}+\frac{2}{15a^2}
-\frac{8A}{3a^2}+A+\frac{8A}{5a^4}\,.
\label{Cv}
\eeq
In the limit \ $M\to 0$ \ we obtain
\beq
\frac{13}{60\,\ep}+k^v_W(a)
\,\,\,\rightarrow \,\,\,\frac{13}{60}\,\Big(
\frac{2}{4-n}-\ga-\ln \frac{\Box}{4\pi \mu^2}\Big)
-\frac{38}{225}\,.
\label{massless V}
\eeq
The divergence and finite constant terms can be canceled by 
local counterterm. The most relevant is the nonlocal term 
with \ $-\frac{13}{60}\,\ln \big({\Box}/{4\pi \mu^2}\big)$,
which is a quantum contribution to the gravitational wave 
equation for the massless limit of the Proca model. The 
corresponding term derived for the gauge vector field 
is just \ $-\frac{1}{5}\,\ln \big({\Box}/{4\pi \mu^2}\big)$. 
The difference between the two coefficients 
\  $1/60 = 13/60-1/5$ \ is the contribution 
of an extra scalar field which was discussed above. In the 
massless limit this field still gives contribution 
to the vacuum effective action, leading to the
discontinuity effect. 

Qualitatively similar situation takes place for the 
massive spin-2 field contributions \cite{duff-discont}. 
However, since such field can be formulated only on special 
backgrounds, there is no real possibility to meet the 
discontinuity in the essential non-local sector. 

\section{Renormalization group for the cosmological 
constant}

\qquad
Perhaps the reader already noticed that the expression
for the effective action of vacuum (\ref{final}) does 
not include nonlocal formfactors for the Einstein-Hilbert 
and cosmological terms. 
As a result we get zero $\be$-functions for the 
corresponding parameters $\beta_{\Lambda}=\beta_{1/G}=0$. 
The reason is that, in the perturbative in $h_{\mu\nu}$ 
approach, we can observe the decoupling for the parameters 
of the higher derivative terms, but not for the cosmological 
and inverse Newton constant. In other words, the physical 
predictions of the $\overline{\rm MS}$ renormalization 
scheme and the ones of the momentum subtraction 
renormalization scheme diverge completely at this point. 
Let us try to explain why this difference takes place. 

\subsection{Once again on the physical renormalization 
group in curved space}

\qquad 
We keep in mind that the renormalization group running 
corresponds, in the UV limit, to the insertion of 
$\,\,\ln(\Box/\mu^2)\,$-like formfactors into the 
effective action. However, as we already discussed in 
the previous sections, such insertions are not possible 
for the CC and $\,1/G$.  

Is it true that physical $\,\be_\La$ and $\be_{1/G}\,$ are 
equal to zero? It is clear that we have no sufficient data 
to claim this. The point is that the corresponding 
$\,\be$-functions come from the contributions of massive 
quantum fields. The consistent formulation of such fields
on curved background require introducing the cosmological 
term into classical action. Then, the 
\ $g_{\mu\nu}=\eta_{\mu\nu}+h_{\mu\nu}$ \ approach implies 
an expansion over the flat background which is not a 
solution of the classical equations of motion. The only 
conclusion we have to make is that the momentum
subtraction scheme and the flat-background expansion are
not appropriate instruments for investigating the 
renormalization group for the cosmological constant. Some 
other methods should be developed before we can arrive at 
some formal results in this area. 

The natural option in this situation is to consider 
the calculations on some special backgrounds. For instance,
the existing methods (e.g., $\ze$-regularization) 
applicable on the de Sitter or anti de Sitter spaces 
\cite{elizalde}, are very 
powerful in sense one can, sometimes, calculate the whole 
effective action. Unfortunately, the information obtained 
in this way is restricted because the curvature is constant 
and all derivatives acting on the curvature are zeros. The 
main disadvantage of the calculations on such backgrounds
is that one can not see the non-localities of the EA.
Furthermore it is not possible to distinguish between 
distinct higher derivative terms from one side and 
the Einstein-Hilbert and cosmological terms from other side.

\subsection{Running cosmological constant}

\qquad 
Since the possible running of the cosmological constant 
is an interesting and ``hot'' issue, one can approach it 
from the phenomenological side and consider applications 
to the ``late'' cosmology. As we have already explained 
in section 2, we shall associate the cosmic scale 
parameter with the Hubble parameter, $\,\mu_c \sim H$ \cite{scale-setting,nova}. 
Let us notice that this identification of the scale has 
certain advantages over other possible choices \cite{reuter}. 

In order to define the possible form of the 
scale-dependence of the quantum corrections to the vacuum 
energy we can recall the considerations presented in 
subsection 3.1 which were based on  covariance. The quantum 
corrections can not be proportional to the odd powers of 
the Hubble parameter. Due to the smallness of 
$\,H_0 \approx 10^{-42}\,GeV$,
compared to the critical density 
$\,\rho_c = \frac{3}{8\pi}H_0^2M_P^2 \approx 10^{-47}\,GeV^4$,
we need to consider only zero-order and second-order in $H$
corrections. 

The zero-order corrections from the particle of the mass 
$\,m\,$ would be proportional to $\,m^4$. It is remarkable 
that the probable value of the neutrino mass 
$\,m_\nu \propto  10^{-12}\,GeV\,$ 
gives the ``quantum correction" to the energy density of 
vacuum $\,\de\rho_\La \sim m_\nu^4\,$
of the right order of magnitude \cite{cosm}. 
The cosmological model based on this form of correction
has been considered in \cite{bauer}.  At the same time 
the hypothesis $\,\de\rho_\La \sim m_\nu^4\,$ has an obvious 
weak point. If the neutrino gives quadratic contribution to 
$\,\rho_\La$, why would not other fermions or bosons give 
such contributions too? There is nothing special in the 
neutrino's vacuum diagrams that could support the idea of
neutrino's exclusive role here. Now, if we assume that 
other massive particles give quadratic contribution, 
the size of such contribution would be huge and destroy 
any reasonable cosmological model.  

As far as the idea of non-decoupling meets serious 
obstacles, let us assume that the Appelquist \& Carazzone-like 
quadratic decoupling holds for the vacuum energy. In this 
case the contribution of the particle of a mass $\,m\,$ 
has the form $\,m^2\mu_c^2$, where $\mu_c \propto H$ is the 
Hubble parameter. It is important to keep in mind that the 
difference between the magnitude of Hubble parameter and 
the one of the masses of {\it all} known particles is 
huge. It is so huge that our intuition concerning 
the smallness of neutrino mass lies here! Let us see. 
We have $\,m/H_0\,$ is $\,10^{-30}\,$ for the 
typical neutrino mass, while in the QCD case we have 
$\,\La_{QCD}/H_0 \sim 10^{-40}\,$ and for the Planck 
mass we have $\,M_P/H_0\sim 10^{-60}$. Hence, qualitatively 
the gap between the cosmic scale $\mu_c$ and neutrino is 
no greater than the one between the  cosmic scale $\mu_c$ 
and the typical QCD scale or between the cosmic scale 
$\mu_c$ and the Planck scale. If we admit the potential 
importance of the neutrino loop or the nonperturbative 
vacuum effects of QCD at the cosmic scale, we have to admit 
that {\it all} existing particles must produce relevant 
contributions. And these contributions must be suppressed
quadratically due to the covariance requirement.  

After all, our considerations lead to the following 
form of quantum correction to the vacuum energy density:
\beq
\,\de\rho_\La \sim \sum \limits_i S_i m^2_i\,H^2\,, 
\label{quadrat}
\eeq
where the sum is over all massive particles, from neutrino 
up to the possible GUT constituents and beyond, to the 
hypothetical Planck-scale particles. The coefficient $S_i$ 
may be different for different particles and, in particular, 
one can expect it to have opposite signs for fermions and 
bosons. After this algebraic summation and adding the 
constant term (required by renormalizability), we arrive 
at the quite general expression for the vacuum energy 
density
\beq
\rho_\La(H)\,=\,\rho_\La^0\,
+ \frac{3\,\nu}{8\pi}\,M_P^2\,\left(H^2-H_0^2\right)\,.
\label{CCH}
\eeq
Here $\rho_\La^0$ is the vacuum energy density in the 
present-day universe, $\rho_\La(H)$ is the vacuum energy 
density and $\nu$ is the unique indefinite component of our 
model. The magnitude and sign of $\nu$ depends, according 
to the eq. (\ref{quadrat}), on the particle's spectrum 
and, mainly, on the spectrum of heaviest particles. Let 
us notice that the formula (\ref{CCH}) holds 
even if we take the higher loop effects into account and 
even if there are some strong nonperturbative effects in 
the unknown high energy part of the particles spectrum. 
In this case the analog 
of the  expression (\ref{quadrat}) will be, of course, 
more complicated due to the possible mass mixing, it will 
depend on the couplings, but the form of (\ref{CCH}) 
will hold, because it is based only on covariance. 

In fact, there is no certainty that the quadratic 
decoupling really takes place. What we can claim is that, 
if there is no cosmic scalar field (or some its 
substitute, like Chaplygin gas etc), the eq. (\ref{CCH}) 
is the unique possible form of the non-constant vacuum 
energy. That is why this form of dependence is worthwhile 
to be explored in details. Imagine we detect someday the 
nonconstant vacuum energy. Then, in order to conclude 
that this indicates some new fundamental physics (it 
may be extra dimensions, quintessence etc) one has to 
rule out the corrections from the quantum matter fields 
on curved background. And these corrections, in the case 
of the vacuum energy, have the form (\ref{CCH}).

It is easy to see that if there 
are no particles beyond the Minimal Standard Model, 
then the product $\,\nu M_P^2\,$ is about 
$M_F^2 \sim 10^5\,GeV^2$. In this case the magnitude of 
the second term in the {\it r.h.s.} of eq. (\ref{CCH}) 
is about $10^{-80}\,GeV^4$ in the recent universe with 
$\,H \approx H_0$, while
$\,\rho_\La^0 \approx 10^{-42}\,GeV^4$. Obviously, this 
means  the quantum contributions are irrelevant and the 
cosmological constant is really a constant. 

On the other hand, if there are particles with the mass 
close to $M_P$, then $\nu$ is of the order one and the 
quantum contribution is of the same order of magnitude 
as $\,\rho_\La^0$. The standard assumption of the gap 
in the mass spectrum between the GUT scale 
$M_X \sim 10^{16}\,GeV$ and $M_P$ produces the value of 
$\nu$ about $10^{-6}$. Of course, the presence of such 
particles does not automatically mean that the value of 
$\nu$ can not be much smaller, because the quantum 
contributions may cancel, e.g., due to supersymmetry.

The cosmological models based on the law (\ref{CCH}) 
have been constructed in \cite{CC-fit} and \cite{Gruni}. 
In both kinds of models the cosmological constant 
terms change together with the Hubble parameter. 
But if the vacuum energy density changes with time, 
this energy has to go somewhere!
The difference between these two models constructed in 
\cite{CC-fit} and \cite{Gruni} is the form 
of the conservation law. In the first case we assume 
the possibility of the energy exchange between the 
vacuum and matter sectors. In the second case such
energy change is forbidden and the conservation law 
is provided due to the $\,H$-dependence of the Newton 
constant $G$. 

The cosmological model of the first sort admits 
analytic solution for the conformal factor \cite{CC-fit}. 
The analysis of density perturbations has been performed 
in the framework of analog model \cite{ana} and also 
through the direct calculus \cite{CCwave}. In both cases 
the obtained limit $\,\left|\nu\right| \leq 10^{-6}$ 
is much stronger than the one which can be expected 
from SNAP \cite{CC-fit}. This bound corresponds to the 
GUT-like particle spectrum. 

The model of the second kind \cite{Gruni} 
is a bit more complicated and the density perturbations 
have not been elaborated yet. However, since at small 
$\,\nu\,$ the two models behave quite similar, one can 
expect that the restrictions for $\nu$ will be also 
similar. It is interesting that this model predicts 
also a logarithmic scale dependence of the Newton 
constant 
\beq
G(H;\nu)=\frac{G_0}{1+\nu\,\log\left(H^2/H_0^2\right)}\,,
\eeq
where $G(H_0)=G_0 = 1/M_P$\,. On the other hand, the 
universality of the effective action of vacuum means that, 
in general, there is similar dependence with $H$ traded 
for the general scale parameter $\mu$.
In particular, in the astrophysical setting this means 
$$
G(r)=\frac{G_0}{1+\nu\,\log\left(r_0^2/r^2\right)}\,,
$$
where $r$ is the distance from the center of the galaxy. 
Then the ``GUT-inspired'' value $\,\nu \sim 10^{-6}\,$ 
is close to the one required for the 
approximately flat rotation curve in the spiral galaxies
\cite{Gruni}. So, the quantum corrections may be 
relevant also for the Dark Matter problem, affecting 
the gravitational law at the galactic scale. This 
distance-dependence does not affect the rotation curves 
inside the Solar System and represents a reasonable
alternative to other models of modified gravitational 
law \cite{Milgrom,Bek}. 

The models developed in \cite{CC-fit} and \cite{Gruni}
show that the time-dependent vacuum energy and also 
distance-depending Newton constant can result from the 
semiclassical theory. However, the existence of the 
quadratic decoupling is hypothetical and needs, at the
first place, a serious theoretical investigation. 
It would be extremely interesting to have a consistent 
QFT model for the effective action behind the possible 
CC running. 
In other words, we need a method for deriving the 
remnant non-local structures in the effective action 
of a massive 
fields. So far, we have no such model. One can prove that
the effective action responsible for the CC running can not be of a 
finite polynomial order in curvatures \cite{CC-fit}, 
but this does not constitute the proof of no-running. 

\section{Do we need Quantum Gravity?}

\qquad
We have started our review from the discussion of Planck 
units, indicating the need of some kind of quantum 
gravity. However, formulating the physical theory 
requires defining the possible area of its application. 
In case of quantum gravity the first choice is the 
region of space-time where the typical energies are 
comparable to the Planck mass. Indeed, this can be 
achieved only in the vicinity of a singularity. 

The singularity is the natural feature of General 
Relativity \cite{Hawk-Pen}, however it is well known 
that the Hawking-Penrose theorem can be violated by 
the semiclassical corrections coming from the quantum
matter fields on classical metric background 
\cite{Ford}. More concrete study has been performed 
in \cite{Ander} for the case of cosmological 
singularity. The theory considered in \cite{Ander}
was based on the conformal anomaly, which we discussed 
in section 4. The result is that the quantum 
corrections which lead to the Starobinsky inflation, 
can erase the initial Big Bang singularity. One can 
call this effect the back reaction of quantum matter 
on the vacuum. Despite the metric is not quantized, 
the form of the vacuum action gets modified due to the 
quantum corrections from matter loops and, 
as a result, in the new theory there is no 
singularity. Similar phenomenon can take place in the 
black hole case. According to the results of 
\cite{FrolVilk}, already the higher derivative terms 
in the classical action of vacuum may switch off the 
singularity in the gravitational collapse. One can 
expect similar effect for the case of anomaly-induced 
effective action, in particular because the metric in 
the interior of the black hole is of the FRW-type
\cite{Frolov}. The situation is expected to be same 
as in the cosmological case \cite{Ander}. 

Finally, let us remark that the Schwarzschild solution
is an essential idealization of the realistic situation 
when the black hole has a nonzero angular momentum. The 
quantum consideration, including the classification of 
the vacuum states of the black hole \cite{balsan}, 
which has been presented in the subsection 5.1, is very 
difficult to generalize for the rotating black hole 
case. Concerning the impact of quantum effects on the
$r=0$ singularity, the geometric structure of the interior 
of the rotating black hole is much more complex than the 
one in a simplified Schwarzschild version. In particular,
for the case of rotating black hole one meets a phenomenon 
which was called ``mass inflation'' \cite{Israel}. The 
description of the possible global structure inside the 
rotating black hole has been given in \cite{FMM} (see 
also \cite{Frolov}).  

Generally speaking,
the results of the existing studies of singularities in 
the theories with semiclassical corrections are not really 
conclusive. In particular, the presence of singularities 
may be affected by angular momentum or the choice of 
initial data in the case of the gravitational collapse 
and, definitely, by the uncertainty which we meet in the 
form of the quantum corrections. 
However, the possibility of a non-singular behavior can 
not be underestimated. If the disappearance of 
singularities in the physically relevant solutions is 
true, the Planck scale physics may become unobservable. 
For instance, as we already mentioned in the Introduction,
the standard argument in favor of the 
quantization of the metric is an inconsistency of the 
semiclassical Einstein equations 
$$
R_{\mu\nu}-\frac12\,R\,g_{\mu\nu}\,=\,8\pi\,G\,T_{\mu\nu}\,.
$$
The \ {\it l.h.s.} \ of 
this equation is classical and deterministic while the  
{\it r.h.s.} depends on quantum variables and therefore
oscillates. However, there is a Planck suppression of 
such oscillations, hence they should be irrelevant at the 
energies much smaller than the Planck ones. Then the 
situation becomes dramatically different if the access
to the Planck energy is really screened by the 
semiclassical modifications of geometry. 

There is an alternative possibility. One may hope to 
observe the effects 
of Quantum Gravity or, e.g., string theory  as small 
effects at low energies. However, similar small effects 
may be, in principle, produced in the framework of the 
semiclassical theory. 

For example, the potentially 
observable time-dependent Dark Energy in the universe 
may be an effect of quintessence or Chaplygin gas, that 
would mean a qualitatively new physics, originating from 
the string theory. Similar effect can be also produced 
by quantum gravity if we accept the popular hypothesis 
on the non-Gaussian fixed point in this theory 
\cite{reuter-RG}. 
However, quite similar effect of a scale-dependent 
vacuum energy may come also from the semiclassical 
gravity \cite{CC-fit,Gruni}. 
The important difference is that, as we stressed in the 
Introduction, the quantum effect of matter fields on 
curved background is something which certainly exists, 
even if we do not know well how to evaluate it. On 
the contrary, we are not certain that there is some 
physical reality behind the string theory or quantum 
gravity. Therefore, the reliable evaluation of the 
semiclassical contributions is a necessary step in the 
study of physical significance of quantum gravity or 
string theory. Before these contributions are 
sufficiently studied, one can not draw conclusions 
about a qualitatively new physics.

\section{Conclusions}

\qquad
The development of experimental and observational 
technologies produce an increasing amount and quality 
of the information about the Universe. For a while, 
the existing data fit with the $\La$CDM model, may be 
with some small modifications like introducing certain 
warmness of the Dark Matter. However,
in the future one can expect to have an access to a more 
detailed description of the expansion of the Universe. It
can not be ruled out that, after all, the Dark Energy will 
be variable \cite{Sahni}. 
This possibility constitutes a strong challenge for the 
theory which has to prepare reasonable explanations of 
this possible variation.  

Does it mean the qualitatively new physics 
which manifests itself in the form of some cosmic field 
or cosmic-scale matter? Such matter would be something 
really unusual, because it must have a very small (if 
nonzero) mass and, 
at the same time, provide self-interaction on the cosmic 
distances. Since nothing like that have been ever seen in 
the laboratories, such cosmic entities would definitely 
mean that we met something beyond the Standard Model or 
its conventional extensions. At the same time, 
one can not rule out that the variable Dark Energy 
is just the result of the quantum effects of matter on 
the classical metric background. Similar situation takes 
place in the study of inflation. The presence of inflaton 
can signal some qualitatively new physics. However
the inflation may be just caused by the Higgs field 
of the Standard Model with the nonminimal coupling,
as it was suggested in \cite{Shaposh}, or result from 
the quantum corrections to the classical vacuum action 
of vacuum in the framework of the Starobinsky model 
\cite{star}, or its modified version \cite{asta}. The 
difference is that 
the vacuum quantum effects or the nonminimally coupled
Higgs are nothing but the manifestations of the 
\ {\it existing} \ physics. According to the conventional  
scientific logic, these options must ``go first'' and 
only if they fail to explain a new phenomena, introducing 
qualitatively new objects must be seriously considered.  

The main obstacle in performing this program is that 
we have no tools to check, for example, whether the 
vacuum quantum effects are really driving the Dark Energy. 
This happens because the relevant part of the vacuum 
effective action comes from the massive fields, it is
non-local and very complicated. In this review we have 
described the present-day situation with quantum 
corrections. As we have seen, the situation is rather 
optimistic for the massless conformal fields. Furthermore, 
we have some models (more or less reliable) for the  
very light quantum fields and, to some extent, for the 
very heavy quantum fields. Unfortunately, we have no 
model for the intermediate case. 

The QFT in curved space-time is about 50 years old. At 
the moment we have a very good understanding of the
renormalization in curved space and of the related 
issues like trace anomaly. Our main hope is to achieve 
better knowledge of the relevant quantum effects in the 
next decades and, in this sense, to match the achievements
of observational cosmology, experimental high energy
physics and phenomenology. 
\vskip 12mm

\noindent
{\large\bf Acknowledgments.} \ 
Author is very grateful to M. Asorey, R. Balbinot, 
I.L. Buchbinder, A. Fabbri, J. Fabris, E. Gorbar, 
G. de Berredo-Peixoto, A. Pelinson and J. Sol\`a for 
numerous discussions and common works and to many other 
colleagues, especially to A.A. Starobinsky and V.P. Frolov, 
for clarifying conversations and correspondence. The present 
review is partially based on the lectures which were given 
at the Universidad de Zaragoza (1995), Centro Brasileiro 
de Pesquisas F\'{\i}sicas 
(1996), Universitat Aut\`onoma de Barcelona (1997),
Universidade Federal de Juiz de Fora (1997 and 2007) and 
at the summer schools ``Quantum Summer" at the Universidade
Federal do Esp\'{\i}rito Santo (1998 and 2007). I am really
thankful to the organizers of these courses and to those 
who were attending lectures and asked questions. The work 
has been partially
supported by the PRONEX project and research grants from
FAPEMIG (MG, Brazil) and CNPq (Brazil), PRONEX project 
from FAPES (ES, Brazil) and CNPq and by the individual 
grants from CNPq and ICTP.


\end{document}